\documentclass[]{bmcart}

\usepackage[utf8]{inputenc} 
\usepackage{graphicx}
\usepackage[ruled,linesnumbered]{algorithm2e}

\usepackage{xcolor,colortbl}
\definecolor{green}{rgb}{0.1,0.1,0.1}

\usepackage{mathtools}
\usepackage{amsfonts}
\usepackage{amsmath}

\usepackage{booktabs}
\usepackage{footnote}
\usepackage{makecell}

\usepackage{tabularx}
\usepackage{times}
\usepackage{caption}
\usepackage{subcaption}

\usepackage{xurl}

\usepackage[]{algorithm2e}
\usepackage{algorithmic}
\usepackage{multicol}

\usepackage{graphicx}

\usepackage{cleveref}
\crefformat{footnote}{#2\footnotemark[#1]#3}

\usepackage{icomma}

\startlocaldefs
\endlocaldefs

\begin{document}

\begin{frontmatter}

\begin{fmbox}
\dochead{Research}


\title{Reaching the bubble may not be enough: news media role in online political polarization}


\author[
   addressref={aff1},                   
   email={jordan@alunos.utfpr.edu.br}   
]{\inits{JK}\fnm{Jordan K} \snm{Kobellarz}}
\author[
   addressref={aff2},
   email={milos.brocic@mail.utoronto.ca}
]{\inits{BM}\fnm{Milos} \snm{Brocic}}
\author[
   addressref={aff1},
   email={@utfpr.edu.br}
]{\inits{AG}\fnm{Alexandre} \snm{Graeml}}
\author[
   addressref={aff2},
   email={dan.silver@utoronto.ca}
]{\inits{DS}\fnm{Daniel} \snm{Silver}}
\author[
   addressref={aff1},
   email={thiagoh@utfpr.edu.br}
]{\inits{TS}\fnm{Thiago H} \snm{Silva}}

\address[id=aff1]{
  \orgname{Informatics, Universidade Tecnológica Federal do Paraná}, 
  \city{Curitiba},                              
  \cny{Brazil}                                    
}
\address[id=aff2]{%
  \orgname{Sociology, University of Toronto},
  \city{Toronto},
  \cny{Canada}
}


\begin{artnotes}
\end{artnotes}

\end{fmbox}


\begin{abstractbox}

\begin{abstract} 

Politics in different countries show diverse degrees of polarization, which tends to be stronger on social media, given how easy it became to connect and engage with like-minded individuals on the web. A way of reducing polarization would be by distributing cross-partisan news among individuals with distinct political orientations, i.e., ``reaching the bubbles''. This study investigates whether this holds in the context of nationwide elections in Brazil and Canada. We collected politics-related tweets shared during the 2018 Brazilian presidential election and the 2019 Canadian federal election. Next, we proposed an updated centrality metric that enables identifying highly central \textit{bubble reachers}, nodes that can distribute content among users with diverging political opinions - a fundamental metric for the proposed study. After that, we analyzed how users engage with news content shared by \textit{bubble reachers}, its source, and its topics, considering its political orientation. Among other results, we found that, even though news media disseminate content that interests different sides of the political spectrum, users tend to engage considerably more with content that aligns with their political orientation, regardless of the topic.
\end{abstract}




\end{abstractbox}
%

\end{frontmatter}


\section{Introduction}

Polarized scenarios are characterized by the tendency of people with similar characteristics, interests, and behaviours to form tightly linked groups \cite{McPherson2001}. In online social networks, this characteristic tends to be stronger because users start to be selectively exposed to similar points of view through mechanisms whose goal is to optimize engagement \cite{pariser_filter_2011}. A small portion of bridging nodes, called brokers, can form weak ties between polarized groups by linking users with access to complementary information and creating opportunities for intergroup contact and exchange of information \cite{burt2003social}. On Twitter's online social network, highly central brokers tend to be related to news media accounts \cite{kobellarz:2019}, through which such media publish links to content available on their respective websites.

This study aims to understand how Twitter users engage with tweets shared by highly central brokers -- users who can reach others with diverging political orientations. For the sake of simplicity, these users are referenced to as \textit{``bubble reachers''} in this study because of their potential ability to reestablish the dialogue among polarized groups of people. More specifically, we study tweets containing URLs to external news by analyzing the relationship between users' political orientation and the content of such news, content source (i.e., website domain), and content topic (extracted using automated processes). We are also interested in understanding the extent to which the exposure of users with a diverging political orientation to the same content contributes to a meaningful engagement, i.e., one that helps to increase the diversity of exchange of ideas, helping to avoid situations such as ``echo chambers'' \cite{mcewan_mediated_2018}. To conduct these analyses, we draw on the 2018 Brazilian presidential election and the 2019 Canadian federal election, two political situations with different polarization levels. For each of these cases, we collected tweets on the weeks before and after voting and the election's results. Datasets are publicly available in: \textit{URL will be released after the review process}. 

The main contributions of this study can be summarized as: (1) The proposition of an updated centrality metric that enables the identification of highly central brokers (\textit{bubble reachers}) that reach users with diverging political opinions; (2) The creation and analysis of tweet engagement networks, which allow us to identify common patterns in Brazilian and Canadian cases, such as: (i) neutral domains tend to distribute information in a balanced way to different polarized groups; however, (ii) when users in these polarized groups are exposed to content that interests both sides of the political spectrum, they tend to engage only with those which align with their own political orientation, regardless of the topic, and (iii) this phenomenon is stronger with the increase in user polarity. Our results indicate that, while news media can ``reach the bubbles,'' i.e., disseminate content that reaches and appeals to different sides of the political spectrum, users still tend to select the information that reinforces their points of view to engage themselves with. This clarifies the nature of news media brokerage, suggesting it can integrate users across the political spectrum, but is less effective in overcoming selective exposure. These findings speak to current debates on media, bias, and polarization.

The rest of this study is organized as follows. Section 2 presents the related work. Section 3 presents the data and methods used. Section 4 presents the results. Finally, Section 5 concludes the study.

\section{Related Works}\label{sec:related}

 Democratic theorists have long extolled the virtues of ties that cross-cut political cleavages \cite{mutz_hearing_2006,putnam_bowling_2000}. Interactions with members of an outgroup are said to promote inclusive relations that encourage civil restraint, and mutual understanding of opposing views \cite{mutz_hearing_2006}. Conversely, in the absence of bridging ties, the public sphere becomes fractured, breeding segmented communities that foment antagonistic partisan identities \cite{putnam_bowling_2000}. Against this backdrop, rising political polarization has been identified as a serious threat facing contemporary democracies. Growing distance between partisans, whether in their attitudes, lifestyles, or relations, has eroded the cross-ties that once tempered partisan conflict \cite{dellaposta_pluralistic_2020}. As a result, partisan politics have become more rancorous, testing the limits of democratic institutions \cite{mason_uncivil_2018}.

Within this context, the role of online media, in particular, has been heavily scrutinized. Early optimism over the democratic potential of networked communities has waned, giving way to concerns over ideologically homophilous networks that proliferate ``fake news'' marred by partisan bias at the expense of factual veracity \cite{faris2017partisanship, pariser_filter_2011}. Some scholars have approached these issues by relying on heuristics such as ``filter bubbles,'' or ``echo chambers'' \cite{mcewan_mediated_2018}. According to these arguments, processes like value homophily and selective exposure that drive partisans to seek like-minded individuals and information corroborating prior attitudes \cite{McPherson2001} are intensified in online spaces, reinforcing ideological viewpoints while insulating partisans from opposing viewpoints \cite{pariser_filter_2011}. Without the tempering effect of this cross-cutting exposure, moreover, some claim partisan commitments become more extreme \cite{mutz_hearing_2006,mason_uncivil_2018}. Consistent with this, research finds that individuals without cross-cutting exposure become less politically tolerant, less likely to regard opposing views as legitimate, and more likely to hold extreme attitudes \cite{huckfeldt_disagreement_2004}.

These developments have brought questions about information brokerage to the fore. Scholars identify brokers as those actors that bridge divides between opposing groups \cite{burt2003social}. Owing to this relational position, brokers benefit from greater control over the transmission of information while also adopting conciliatory attitudes that mediate tensions \cite{burt2003social}. As gatekeepers of political communication, media outlets have traditionally assumed this role \cite{graber_mass_2018}. By selecting the news that citizens read and scrutinizing the quality of information they consume, media outlets integrate citizens into common narratives \cite{anderson_imagined_1991}. Some scholars refer to this as ‘agenda-setting,’ whereby news media have the power to integrate public opinion by telling them what the important issues, worthy of attention, are \cite{mccombs2020setting,lippmann2017public}.

The emergence of social media, however, is said to have upended this role \cite{ferguson_square_2017}. Platforms like Twitter have reconfigured how information is diffused by enabling grassroots actors to bypass traditional media outlets and spread crowdsourced information \cite{ferguson_square_2017}. As discussed, these platforms are seen as encouraging dynamics like homophily, moral sensationalism, and ``factual polarization'' that drive echo chambers \cite{brady_emotion_2017}. Taking heed of these issues, social media platforms and mainstream news outlets have sought to remedy this situation by re-asserting their brokerage function. Initiatives include expanding ``fact-checking'' efforts, with social media platforms like Twitter and Facebook collaborating with third-party programs that independently review content and flag false information. By directing citizens to institutionally validated facts, such initiatives intend to quell polarization by offering a ``neutral'' medium that foregrounds partisan debate and counteracts ideological bubbles. 

Recent research, however, suggests that the ideological bubble perspective misdiagnoses the nature of online polarization in certain ways \cite{judith_moller_filter_2021}. Empirical evidence finds that online ideological homophily is overstated, demonstrating that online networks are diverse in viewpoints and news sources, often more so than offline networks \cite{beam_facebook_2018}. In fact, users often seek out counter-attitudinal political information. Morales et al. \cite{de_francisci_morales_no_2021} find that during discussions of the 2016 US elections, \textit{Reddit} users were less likely to engage with members of their own party than they were to engage across party lines. Likewise, Dylko et al. \cite{dylko_impact_2018} find that online engagement is higher when users are exposed to counter-attitudinal items. Rather than moderating viewpoints, however, exposure to counter-attitudinal information can backfire, reinforcing commitment to prior beliefs \cite{bail2018exposure}. Summarizing the empirical record, Moller \cite{judith_moller_filter_2021} claims that ``the reality of social media and algorithmic filter systems today seems to be that they inject diversity rather than reducing it''. Despite the normative benefits typically associated with diverse social ties \cite{mutz_hearing_2006}, however, it appears that exposure to diverse content on digital spaces can sometimes worsen polarization \cite{stray_2021}. Overall, this research challenges the premise that cross-cutting ties mediate conflict in online settings, suggesting that exposure to opposing views alone appears insufficient as a remedy to polarization. The information we receive and the evaluations we make appear to be guided by an effort to preserve the integrity of prior beliefs, even when confronted with opposing evidence \cite{jerit_partisan_2012}. Brokers who reach ideological bubbles by affording universal access to similar content might not necessarily quell polarization, as partisan bias continues to motivate users to select stories and narratives that corroborate prior beliefs, and reconcile challenging information by rejecting its validity \cite{mcewan_mediated_2018}. 

Brokers’ ability to temper opposing positions may thus be limited because of selective exposure. That said, they may still mediate polarization in other ways. As mentioned above, an important function of brokerage traditionally provided by news media involves ‘agenda-setting’, whereby news media effectively tell people what topics to pay attention to, even if they don’t dictate what the specific positions should be  \cite{mccombs2020setting,lippmann2017public}. From that angle, established news media can still promote integration by delineating important topics of public discussion, thus drawing interest from partisans across the spectrum, uniting them in their attention, even if as hostile counterparts. Given the fragmentation of online media, however, the extent to which established news media that aim to be neutral brokers succeed in this capacity is unclear.

Taken together, this research raises numerous questions on brokerage in online settings. Our study seeks to adjudicate divergent perspectives by investigating how online brokers mediate polarization during two elections in Brazil and Canada. Specifically, we pursue four research questions: (1) How can we identify social media brokers (\textit{bubble reachers}) who reach users across partisan divides? (2) Does content shared by \textit{bubble reachers} elicit high or low levels of polarization? (3) Are more ``neutral'' users less likely to engage with polarized content when compared to ``polarized'' users? (4) Does the source of content shared by \textit{bubble reachers} affect the degree to which it elicits a polarized response, based on the specific article or its semantic content (represented by its topic)?

\section{Data and Methods}\label{sec:method}

\subsection{Data}

We collected tweets from the Twitter streaming API regarding the 2018 Brazilian presidential election and the 2019 Canadian federal election. To obtain tweets related to the Brazilian and Canadian cases, we collected trending political hashtags ranked by Twitter on those countries on different moments of the preceding collection day to use as filtering criteria (``seeds'') on the Twitter API. We made sure to have a balanced number of trending hashtags supporting different candidates. Examples of hashtags for the Brazilian scenario are: \#bolsonaro, \#haddad, \#elenao, \#elesim, \#lula, \#TodosComJair, and for the Canadian scenario: \#CanadaElection2019, \#cdnpoli, \#TrudeauMustGo, \#TrudeauWorstPM, \#TeamTrudeau, \#Trudeau4MoreYears. Data collection for Brazil started $6$ days before the first-round vote, from 2018-Oct-01, and ended $15$ days after the second-round vote, on 2018-Dec-11. This dataset was filtered to obtain only tweets whose language of the text and the author's profile was Portuguese, resulting in $43,140,298$ tweets. For the Canadian case, tweets' collection started $34$ days before the polls on 2019-Sep-18 and ended $16$ days after the polls on 2019-Sep-06. This dataset was filtered to obtain only tweets whose language of the text was English or French, resulting in $12,426,349$ tweets. We did not filter by users' profile language for the Canadian dataset, because the Twitter Streaming API was no longer making this attribute available on tweets body. Both Brazilian and Canadian datasets were analyzed separately. 

The political situation for each election provides important context for the analysis. In Brazil, during the election time, the polls\footnote{\url{http://media.folha.uol.com.br/datafolha/2018/10/19/692a4086c399805ae503454cf8cd0d36IV.pdf}} indicated high polarization among voters. The dispute was between Jair Messias Bolsonaro, representing the possibility of a 15-year break from the ruling Workers Party (PT), and Fernando Haddad, representing the continuity of PT’s rule, after a brief period in which a PT president elected was replaced by her vice-president as a result of impeachment. The election culminated in Bolsonaro's victory, with $55.13\%$ against $44.87\%$ of the valid votes in favour of Haddad\footnote{\url{https://politica.estadao.com.br/eleicoes/2018/cobertura-votacao-apuracao/segundo-turno}}. In Canada, Justin Trudeau represented the Liberals, which had previously held a parliamentary majority after unseating the Conservatives in 2015. In a close election, the Liberals won 157 ($39.47\%$) seats in parliament, while the Conservatives, led by Andrew Scheer, won 121 ($31.89\%$)\footnote{\url{https://newsinteractives.cbc.ca/elections/federal/2019/results}}. The (left-wing) New Democratic Party continued to lose ground from its 2011 peak, especially in French Quebec, where the Liberals and the separatist Bloc Québécois subsequently gained ground. The 2019 election resulted in the Liberals forming a minority government, which have historically exhibited instability, since the prime minister relies on representatives of other parties to remain in power\footnote{\url{https://web.archive.org/web/20130627154515/http://www2.parl.gc.ca/Parlinfo/compilations/parliament/DurationMinorityGovernment.aspx}}. Despite the multi-party character of the Canadian political system, at least since the 1980s, national politics has largely revolved around left-right differences \cite{cochrane2015left}.

\subsection{User Polarity Estimation}
\label{subsection:user-polarity-estimation}

The next step was to estimate each user's polarity based on the retweeted content. For this, we used the political orientation of the hashtags that users applied in their tweets as an indication of their polarity. For this task, we identified $87,620$ and $86,959$ unique hashtags for the Brazilian and Canadian cases, respectively. For each of these two sets, we extracted the top 100 most frequent hashtags and classified them manually according to their political orientation, ``$L$'', ``$N$'', ``$R$'' and ``$?$'', were used to represent left-wing, neutral, right-wing and uncertain political leaning, respectively. Six volunteers (not the authors), three in each country, helped to classify all the top 100 hashtags without interference from one another. We maintained only the hashtags whose classification was the same for the three raters, resulting in 64 and 78 out of 100 hashtags for Brazil and Canada, respectively. We relied on the Fleiss' kappa assessment \cite{fleiss1981measurement} to measure the agreement degree between the raters, obtaining $\kappa = 0.63$ for Brazil, which means ``substantial agreement'', and $\kappa = 0.80$ for Canada, meaning ``almost perfect agreement'', according to Landis and Koch's (1977) \cite{landis1977measurement} interpretation of kappa scores. 

To classify the rest of the hashtags at scale, we assumed that hashtags describing a common topic usually occur together in the same tweet. This way, a network of hashtag co-occurrences was built for Brazil and Canada, in which a node represented a hashtag and an edge between two nodes, the occurrence of both in the same tweet. Standalone hashtags were eliminated. A semi-supervised machine learning algorithm \cite{zhu2003semi}, which uses the edges' weight to compute the similarity between nodes, was applied over the network to label unlabeled hashtags starting from the manually classified hashtags. By applying this procedure, $57,487$ ($65.6$\%) and $67,639$ ($77.8$\%) hashtags were classified in the Brazilian and Canadian datasets, respectively. For the Brazilian case we obtained 35,788 (62.3\%) hashtags classified as ``$R$'' (right), 643 (1.1\%) as ``$N$'' (neutral), 21,039 (36.6\%) as ``$L$'' (left), and 17 (0.0\%) as ``$?$'' (uncertain). For the Canadian case we obtained 8,963 (13.3\%) as ``$R$'', 52,416 (77.5\%) as ``$N$'', 2,597 (3.8\%) as ``$L$'', and 3,663 (5.4\%) as ``$?$''. To assess the consistency of this method, the same hashtag classification procedure was performed $100$ times, but randomly hiding 10\% of the manually classified hashtags each time. Classification results were submitted to the Fleiss' kappa assessment \cite{fleiss1981measurement} (simulating $100$ raters), where we obtained a $\kappa = 0.84$ for Brazil, and $\kappa = 0.84$ for Canada, meaning ``almost perfect agreement'' in both cases. We also considered a more aggressive strategy, hiding 20\% of the manually classified hashtags, obtaining $\kappa = 0.78$ for Brazil, and $\kappa = 0.73$ for Canada, meaning ``substantial agreement'' in both cases, showing that the classification procedure was robust.

Results showed an imbalance in both datasets between hashtags classified as ``$N$'' and those classified as ``$R$'' or ``$L$''. Taking a close look at the Canadian dataset, it is possible to conclude that users tend to apply neutral hashtags more frequently, for example, \#cdnpoli, \#elxn43, \#onpoli, and \#abpoli, together with polarized hashtags, such as \#TrudeauMustGo, \#blackface, \#ChooseForward, or \#IStandWithTrudeau, which, in turn, are applied in a proportionally smaller number, helping to explain the more prominent number of neutral hashtags in this case. On the other hand, in the Brazilian dataset, the use of left and/or right hashtags is higher than neutral hashtags, even when they appear together. This imbalance on both datasets was addressed by adding weights to each class during the users' polarity estimation, presented in Equation~\ref{eq_PH}. Given the low relevance of hashtags with uncertain positioning ``$?$'' for the analysis, only those classified as ``$L$'', ``$N$'' and ``$R$'' were maintained. Finally, we removed all tweets from the dataset that did not contain any classified hashtag, to preserve only political domain-related tweets. This removal resulted in $4,217,070$ ($6,1$\%) and $2,304,911$ ($17,4$\%) tweets for the Brazilian and Canadian datasets, respectively. 

The next step was to classify users according to their political orientation, which was performed on a weekly basis. For this, we separated each dataset into six weeks, starting on Monday and ending on Sunday. Less active users, with less than five tweets per week, were removed. This was done because these users did not create enough tweets to estimate their polarity. With that, we obtained $72,576$ and $26,815$ unique users for the Brazilian and Canadian datasets, respectively. For each of these users, a list of hashtags used in all of their tweets for the week was created, and their polarity ($P(H)$), calculated using Equation \ref{eq_PH}:

\begin{equation}
    P(H)=\frac{|H_R|\times W_R-|H_L| \times W_L}{|H_L| \times W_L+|H_N| \times W_N+|H_R| \times W_R},
  \label{eq_PH}
\end{equation}

where $H_L$, $H_N$ and $H_R$ are the hashtag multisets (a set that allows for multiple instances for each of its elements) for classes ``$L$'', ``$N$'' and ``$R$'', respectively. $ W_L=avg(|H_N|, |H_R|)/|H|$, $ W_N=avg(|H_R|, |H_L|)/|H|$, and $W_R=avg(|H_L|, |H_N|)/|H|$ are the weights for classes ``$L$'', ``$N$'' and ``$R$'', respectively. In these equations, $avg(.)$ is a function that returns the average number of hashtags in two sets, and $H$ is a set containing all hashtags applied by a user, i.e., $H = H_L \cup H_N \cup H_R$. These weights are important to mitigate the class imbalance characteristic of our datasets. This is inspired by usual tasks in classification scenarios with imbalanced classes \cite{tan2016introduction}. The general idea is to penalize classes with a higher number of hashtags, such as the case of class ``N'' in Canada, and ``R'' and ``L'' in Brazil, and increase the relevance of classes with fewer hashtags, such as `` R '' and ``L'' in Canada, and ``N'' in Brazil. This is necessary to avoid polarity estimation biased to the most frequent hashtags users’ tweeted. Without this strategy, most users would have their polarity estimated wrongly as neutral in the Canadian case. For example, a user who tweeted \#ChooseForward, \#ChooseForwardWithTrudeau, \#cdnpoli, \#elxn43, \#ScheerCowardice, \#onpoli, \#VoteLiberal, \#cdnpoli, \#elxn43, \#ItsOurVote, \#CPC and \#climatestrikecanada, which is a representative example of our dataset, would have a P(H) value equal to $-0.1$ without using weights, and a P(H) value equal to $-0.5$ with weights, better reflecting the polarity of the user. Similarly, a user who applied the hashtags \#cdnpoli, \#elxn43, \#elxn43, \#RCMP, \#cdnpoli, \#TrudeauIsARacist and \#TrudeauCorruption, would have a P(H) value of $+0.3$ without weights and equal to $+0.7$ with weights, again the use of weights expresses a user's polarity much better. Without weights, the opposite would happen in the Brazilian case; most users would have their P(H) values skewed to extremes improperly.

The result of $P(H)$ is a value in the continuous range $[-1.0; +1.0]$. Positive values represent a right-wing political orientation, negative values represent a left-wing orientation, and values close to $0.0$ represent a neutral orientation. Based on this result, we labelled users according to their political orientation: the first third of values on the P(H) scale represents the left-wing (L) users, with $P(H) \in [-1; -1/3[$, the second third, the neutral users (N), with $P(H) \in [-1/3; 1/3]$ and the last third, the right-wing users (R), with $P(H) \in ]1/3; 1]$. It is important to note that the use of the terms left ($L$), right ($R$) and neutral ($N$) to denote the political orientation of hashtags and users was a simplification used to make it possible to compare the two political situations through a common and simplified categorization \cite{cochrane2015left}. 

\subsection{Detecting  \textit{bubble reachers}}

After identifying all users' political orientation, a network was created for each week, connecting one user to another through retweets. Each network was represented in the form of a weighted undirected graph, where each user is a node, and the retweet is an edge that starts from the user who was retweeted and ends at the user who retweeted. The network is undirected because the incoming activity of any individual node -- how much they retweet -- is dwarfed by the outgoing activity of popular nodes -- how much they get retweeted--, thus minimizing problems due to this characteristic. The edge weight represents the number of retweets between users. Self-loops and isolated nodes were eliminated. 

To achieve the main goal of our study, we needed to detect highly central nodes linked to both sides of the network, i.e., we needed to detect brokers (\textit{bubble reachers}) on the network capable of reaching users with diverging political views. For this task, \textit{betweenness} centrality \cite{freeman1978centrality} could be a metric to be applied as a starting point, because it ranks nodes by their capacity of acting as information bridges between more pairs of nodes in the network, relying on the number of shortest paths that pass through each node. However, in a polarized situation, we can have nodes with a high \textit{betweenness} degree within and between polarized groups. That is, highly influential nodes inside bubbles could be ranked the same as highly influential nodes between bubbles. The former are called ``local bridges'' and the latter ``global bridges'' \cite{jensen2016detecting}. We were interested only in the global bridges, nodes that most of the time act as brokers, by linking both sides of the network. 

Considering that \textit{betweenness} centrality was not an appropriate metric to distinguish local bridges from global bridges, we identified in the literature a relatively new centrality metric called ``bridgenness'' \cite{jensen2016detecting}. This metric is based on the \textit{betweenness} algorithm, but while computing the number of shortest paths between all pairs of nodes that pass through a source node, it does not include the shortest paths that either start or end at the immediate neighborhood of the source node \cite{jensen2016detecting}. Even though the \textit{bridgeness} algorithm could better emphasize global than local bridges, it brings up a problem when the considered network has a small average path length, which is precisely what happens with the retweet networks we are analyzing. This is because it could disregard some important small paths that either start or end on the neighborhood of a node. Considering that we already know the political orientation of all users in our dataset and that users with similar orientations tend to form tightly linked groups on the network, we used the political orientation as a filtering criterion for the shortest paths.

Algorithm~\ref{alg:bubble-popper} presents this proposed process, which we called the ``\textit{intergroup bridging}'' algorithm. This algorithm builds on the \textit{betweenness} and \textit{bridgeness} algorithms. Still, while computing the number of shortest paths between all pairs of nodes that pass through a node, it does not include the shortest paths that either start or end at the immediate neighborhood of a certain node if the considered node on the neighborhood has the same class (political orientation, in our case) as the source node -- a key twist added to the \textit{bridgeness} algorithm is this restriction to be from a different class. Put in simple words: this new algorithm measures a node's capacity to disseminate information to distinct groups on the network, with a different political orientation from itself. To construct the \textit{intergroup bridging} algorithm, we relied on the Brandes ``faster algorithm'' \cite{brandes2001faster}. These proposed changes are presented in line~\ref{alg:line-36} -- in the count of the shortest paths that pass through node $w$, it is verified if the considered path is not a self-loop with $w \neq s$ (which already exists in the original Brandes' algorithm) and if $s$ is not in the neighborhood of $w$, with $A[w, s] == 0$ or, if $s$ is in the neighborhood of $w$, with $A[w,s] >= 1$ and $s$ has a different political orientation from $w$, with $O[w] \neq O[s]$. We refer to the measure created by this algorithm as ``\textit{intergroup bridging} centrality.'' Note that this metric is also applicable to other problems with the same characteristics. 

\begin{algorithm}[tb]

\begin{algorithmic}[1]
    \STATE $A[u,v] \gets $ adjacency matrix\;
    \STATE $O[v] \gets $ node orientation, $v \in V$\;
    \STATE $C_B[v] \gets 0, v \in V$\;
    \FOR{$s \in V$}
        \STATE $S \gets $ empty stack\;
        \STATE $P[w] \gets $ empty list, $w \in V$\;
        \STATE $\sigma[t] \gets 0, t \in V;  \sigma[s] \gets 1$\;
        \STATE $d[t] \gets -1, t \in V;  d[s] \gets 0$\;
        \STATE $Q \gets $ empty queue\;
        \STATE enqueue $s \to Q$\;
        \WHILE{$Q$ not empty}
            \STATE dequeue $v \gets Q$\;
            \STATE push $v \to S$\;
            \FORALL{neighbor $w$ of $v$}
                \STATE // $w$ found for the first time?
                \IF{$d[w] < 0$}
                    \STATE enqueue $w \to Q$\;
                    \STATE $d[w] \gets d[v] + 1$\;
                \ENDIF
                \STATE // shortest path to $w$ via $v$?
                \IF{$d[w] = d[v] + 1$}
                    \STATE $\sigma[w] \gets \sigma[w] + \sigma[v]$\;
                    \STATE append $v \to P[w]$\;
                \ENDIF
            \ENDFOR
        \ENDWHILE
        \STATE $\delta[v] \gets 0, v \in V$\;
        \STATE // $S$ returns vertices in order of non-increasing distance from $s$
        \WHILE{$S$ not empty}
            \STATE pop $w \gets S$\;
            \FOR{$v \in P[w]$}
                \STATE $\delta[v] \gets \delta[v] + \frac{\delta[v]}{\delta[w]} \cdot (1 + \delta[w])$\;
            \ENDFOR
            \STATE // this is the part where \textit{intergroup bridging} differs from betweenness and bridgeness algorithms 
            \IF{$w \neq s$ and ($A[w,s] == 0$ or $A[w,s] >= 1$ and $O[w] \neq O[s]$)} \label{alg:line-36}
                \STATE $C_B[w] \gets C_B[w] + \delta[w]$
            \ENDIF
        \ENDWHILE
    \ENDFOR
\end{algorithmic}
\caption{\textit{intergroup bridging} algorithm, adapted from Brandes' ``faster algorithm'' for \textit{betweenness} centrality \cite{brandes2001faster} and \textit{bridgeness} algorithm \cite{jensen2016detecting}.}
\label{alg:bubble-popper}
\end{algorithm}

\begin{figure}[ht!]
    \centering
    \begin{subfigure}{12cm}
        \centering
        \includegraphics[width=8cm]{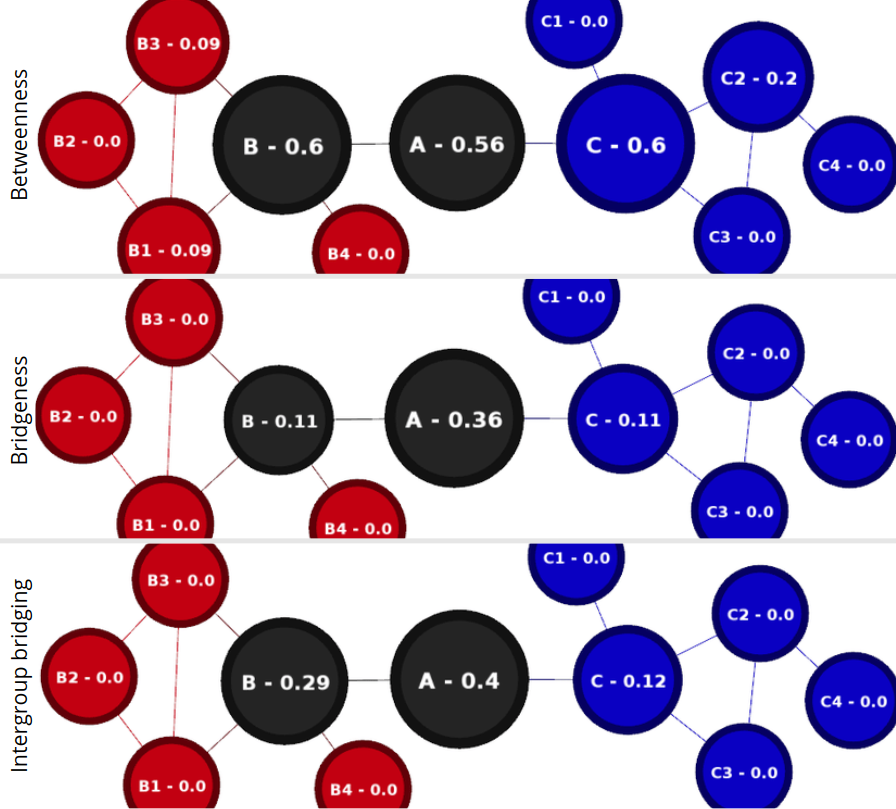}
        \caption{Synthetic network.} 
    \end{subfigure}%
    \\
    \begin{subfigure}{12cm}
        \centering
        \includegraphics[width=12cm]{"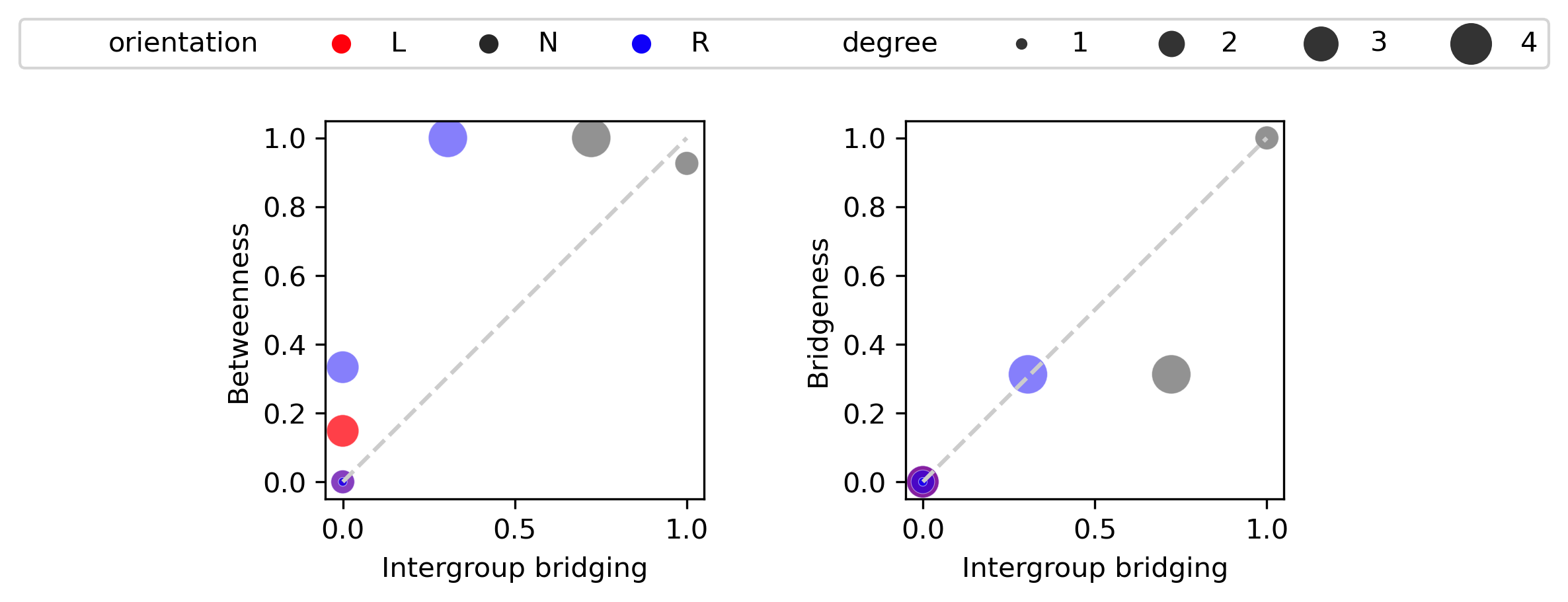"}
        \caption{Centralities relationship.}
    \end{subfigure}
    \caption{Centralities comparison on a synthetic network.}
    \label{fig:centralities_syntethic}
\end{figure}

\begin{figure}[ht!]
    \centering
    \centerline{\includegraphics[width=0.99\textwidth]{"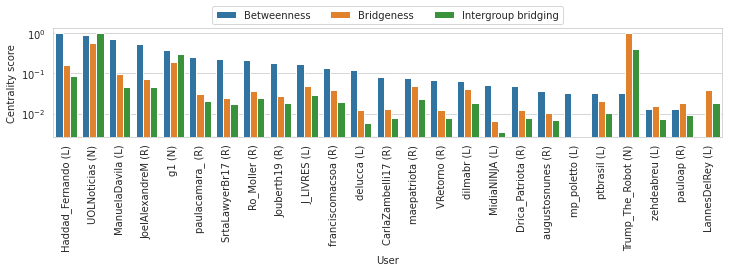"}}
    \caption{Centrality comparison among top 25 users according to \textit{betweenness} for the Brazilian dataset - week 4.}
    \label{fig:top_centrality_comparison}
\end{figure}

To illustrate the difference between centrality metrics, a synthetic network was used, the same one evaluated by Jensen et al. (2016) \cite{jensen2016detecting} to compare the metrics of \textit{betweenness} and \textit{bridgeness}, but with the addition of a label for each node representing its political orientation (L, N or R), to allow the \textit{intergroup bridging} metric computation. Figure~\ref{fig:centralities_syntethic} presents this network on (a), including the computed values for all metrics, with colour-coded nodes according to political orientation, with L in red, N dark-coloured, and R in blue, and two scatter plots on (b), one representing the relationship between \textit{intergroup bridging} with betweenness, and another between \textit{intergroup bridging} with bridgeness, where the colors reflect the user political orientation, and the size of the point its degree (the bigger, the higher), with all values normalized in the $[0.0;1.0]$ interval by each metric, to allow a fair comparison. In this figure, it is possible to observe that the \textit{intergroup bridging} metric was more efficient in detecting nodes that bridge distinct groups on the network with a different political orientation from itself, represented by nodes A and B. To complement this analysis, Figure~\ref{fig:top_centrality_comparison} shows centrality values for the $25$ users with the highest \textit{betweenness} values in Brazil in week $4$. These values were normalized using a min-max strategy for each metric. The user's political orientation (L, N, or R) is presented next to their name. It is possible to note that the values of \textit{bridgeness} and \textit{intergroup bridging} centrality follow a similar pattern for most users in Brazil, being the former metric slightly higher than the latter, showing that those metrics capture similar information for most cases. However, they differ considerably in specific cases, especially when important nodes reach different sides of the spectrum in a more balanced way, such as the case for @UOLNoticias and @g1, two Brazilian news media profiles.

\begin{figure}[ht!]
    \includegraphics[height=8cm]{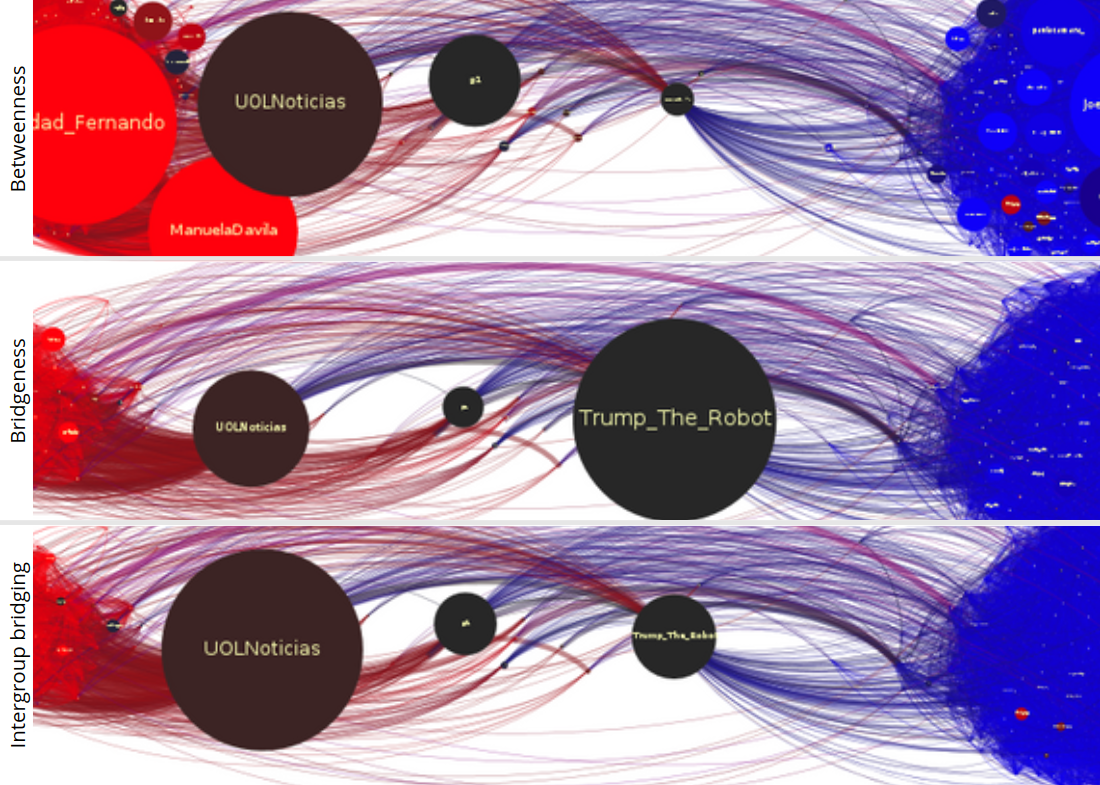}
    \caption{Comparison among centralities on week 4 in Brazil.} 
    \label{fig:centralities}
\end{figure}

This result can be difficult to understand without being familiar with the network structure; thus, Figure~\ref{fig:centralities} compares the same metrics on a zoomed example of the same network analyzed above. Nodes are sized by their respective centrality measure. On the left and right corner of the figures, one can view the left-wing (red) and right-wing (blue) groups. Between them, a small number of nodes (dark-coloured) link these groups. On the first representation for \textit{betweenness} centrality, two big nodes appear. These are for accounts @Haddad\_Fernando and @ManuelaDavila, representing Fernando Haddad and Manuela Davila, candidates for presidency and vice-presidency. On the left-wing side of the figure, these accounts were ranked higher than the bridging nodes @UOLNoticias and @G1. On the second representation (\textit{bridgeness} centrality), as expected, nodes from polarized groups were also not highlighted. Rather, @Trump\_The\_Robot, a spamming account banned from Twitter after data collection, was ranked higher than @UOLNoticias and @G1. And finally, on the third representation for \textit{intergroup bridging} centrality, nodes from polarized groups also were not highlighted. However, in contrast to \textit{bridgeness} centrality, @UOLNoticias ranked higher than @Trump\_The\_Robot and @G1. This same pattern prevailed in other weeks. For Canada (not shown due to lack of space), the \textit{intergroup bridging} also helps to highlight important nodes that receive the attention of users from different sides of the political spectrum. These findings regarding the effectiveness of the \textit{intergroup bridging} centrality algorithm speak to our first research question.

\subsection{Domain, Content and Topic Polarity Estimation}
\label{subsection:domain-content-topic}

The next step was to extract entities (i.e., domain, content, and topic) related to the links to external sites present in tweets made by  \textit{bubble reachers}. Recall that content refers to news represented by its URL, domain refers to the news website domain, and the topic refers to the latent topic in the content extracted using standard automated processes. The process for extracting and estimating the polarity of these entities is presented in this section.

\textit{Intergroup bridging} centrality was used to obtain the $100$ users that had the highest value for this metric in each of the weeks. Knowing the most central accounts, we obtained all retweets made by other users from content shared by this specific group of highly central accounts that included links to external sites, except social networking sites (i.e., Facebook, Instagram, Linkedin, Reddit, Periscope, and Snapchat) and content aggregators (i.e., YouTube, Google News, Apple Podcasts and Globo Play - in case of Brazil). In total, $299,873$ and $261,703$ retweets with links were identified in the Brazilian and Canadian datasets, respectively. Shortened links have been transformed to its original form, and those whose related domain was retweeted less than $10$ times or had only a single link shared along all weeks were removed. As a result, only links from domains with at least two distinct links shared during the data collection period were maintained. To capture the textual content and metadata from pages referred to in these links, the \textit{news-please}\footnote{\url{https://pypi.org/project/news-please}} library was used. A total of $ 370 $ and $ 625 $ unique links were extracted from the Brazilian and Canadian datasets, respectively. Any textual content with less than $300$ characters was removed to avoid potentially incomplete articles, resulting in $ 338 $ and $ 484 $ unique contents in the Brazilian and Canadian datasets, respectively. Then, the textual content was pre-processed in the following order: (1) removal of stop-words from specific lists for the languages of each dataset using the NLTK library\footnote{\url{https://www.nltk.org}}; (2) removal of tokens that were not nouns, proper names, adjectives or verbs using specific pre-trained POS-tagging models for the languages of each dataset using the Spacy library\footnote{\url{https://spacy.io.}}; (3) stemming of tokens using pre-trained models for the languages of each country through the Spacy library; (4) transformation of tokens into lower case and removal of special characters and accents; (6) extraction of bi and trigrams using the Gensim library\footnote{\label{note:radimrehurek}\url{https://radimrehurek.com/gensim}}.

Having cleaned the data, we extracted topics from the pre-processed textual content. For this task, we applied the Latent Dirichlet Allocation (LDA) \cite{blei2003latent} algorithm using the Gensim library implementation. This algorithm allows for identifying topics in a set of texts, considering that each text has a mixture of topics \cite{blei2003latent}. 
The choice of the number of topics needs to be made manually for the LDA algorithm. Therefore, multiple models were created for each dataset with values for the number of topics in the range from 1 to 50. Within this range, we identified the model whose quantity of topics had the highest degree of semantic similarity between texts through the coherence metric generated with the $C_v$ model \cref{note:radimrehurek}, which is based on the indirect cosine similarity of most representative words from each topic throughout all documents on the dataset. For all cases, the same seed was used so that the results of identifying topics could be replicated. Using this method, we found 48 topics in Brazil and 26 in Canada. After extracting topics with this method, it was analyzed the topic dominance of each content, from which it was found that 87\% and 68\% of contents in the Brazilian and Canadian datasets, respectively, were dominated by only one topic, with a dominance of at least 80\% over other topics. Considering this result, we extracted the dominant topic of each content and checked the mean number of contents tied to each dominant topic. In the Brazilian case, it was found a mean value of 6.5 contents per dominant topic, with a standard deviation of 2.5, and, in the Canadian case, a mean of 15.0 contents per dominant topic, with a standard deviation of 4.1. These results indicate that most topics were well-defined (less nebulous) and evenly distributed between contents. Finally, the authors manually evaluated all topics identified and reached a consensus on whether the topics were closely related to the respective political situations.

The last step was to estimate the polarity of domains, content, and topics based on the polarity ($P(H)$) of the users who retweeted a tweet. For this task, a new metric called relative polarity ($RP(H)$) was created, which is calculated as follows: (1) for each entity (i.e. domain, content or topic), a list with 21 positions was created, where each cell computes the number of retweets that the entity received from users in the polarity bins represented by the set \{$ -1.0, -0.9, ..., 0.0, ..., +0.9, + 1.0 $\}; (2) the entities were allocated in a matrix, in which each row represents an entity and each column one of the 21 polarity bins; (3) in order to avoid data imbalance for each bin, considering the overall dataset, each cell of the matrix was divided by their respective column maximum value; (4) for each entity (row), we normalize the values using a min-max strategy, putting the values on a $[0.0, 1.0]$ interval; (5) for each entity (row), we summed each cell value multiplied by its respective polarity, for example, the first cell value was multiplied by $-1.0$, the second cell by $-0.9$, the third by $-0.8$, and so on, until $+1.0$. If the cell value was equal to zero or the cell represented the polarity $0.0$ it was disregarded and not counted; (6) finally, the sum result was divided by the number of considered cells, which became the RP(H) value for the entity. The resulting value of the metric RP(H) was interpreted in the same way as the metric P(H), presented in Subsection~\ref{subsection:user-polarity-estimation}.

\section{Results}\label{sec:results}

\subsection{Polarization analysis}

\begin{figure}[ht!]
    \centering
    \begin{subfigure}{6cm}
        \centering
        \includegraphics[width=5cm]{"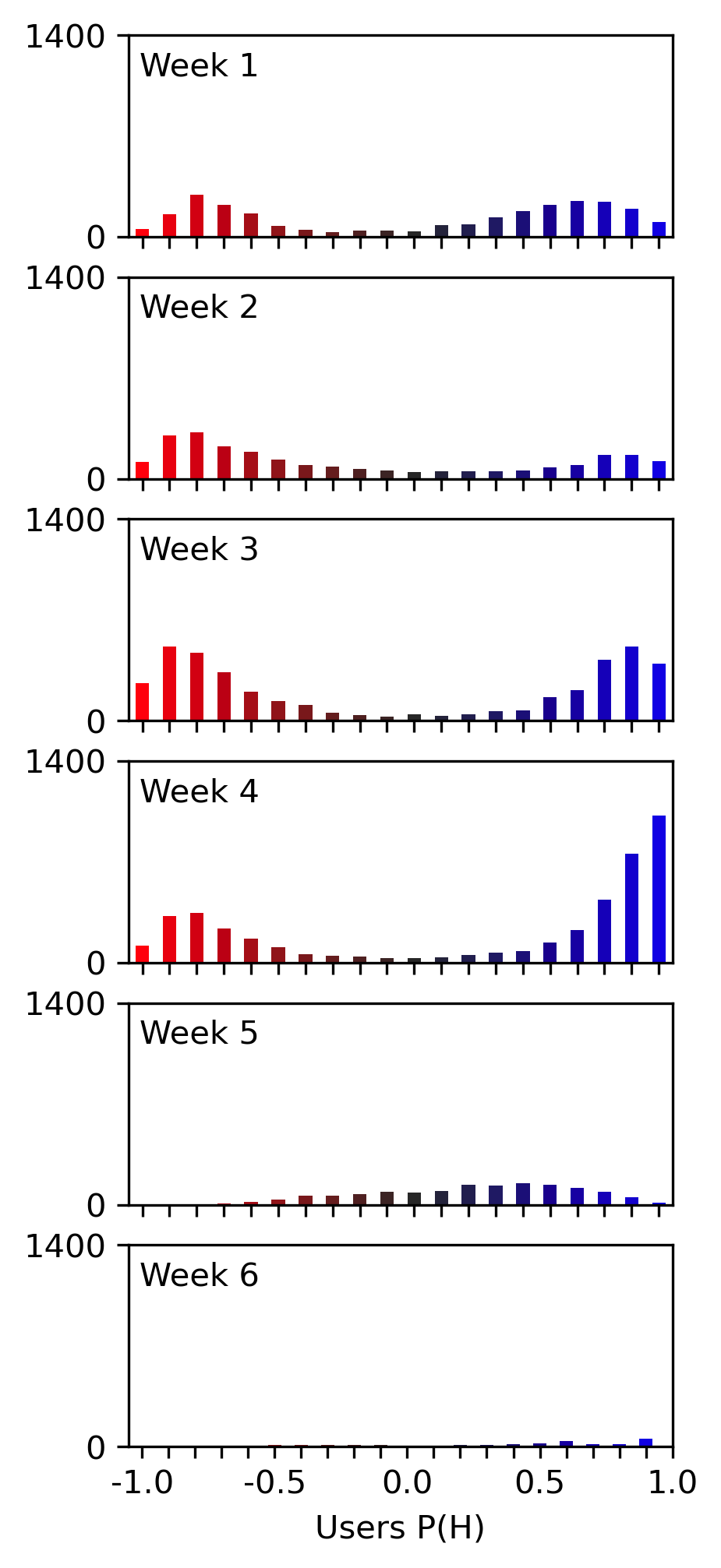"}
        \caption{Brazil}
        \label{fig:polarity-bra}
    \end{subfigure}%
    \begin{subfigure}{6cm}
        \centering
        \includegraphics[width=5cm]{"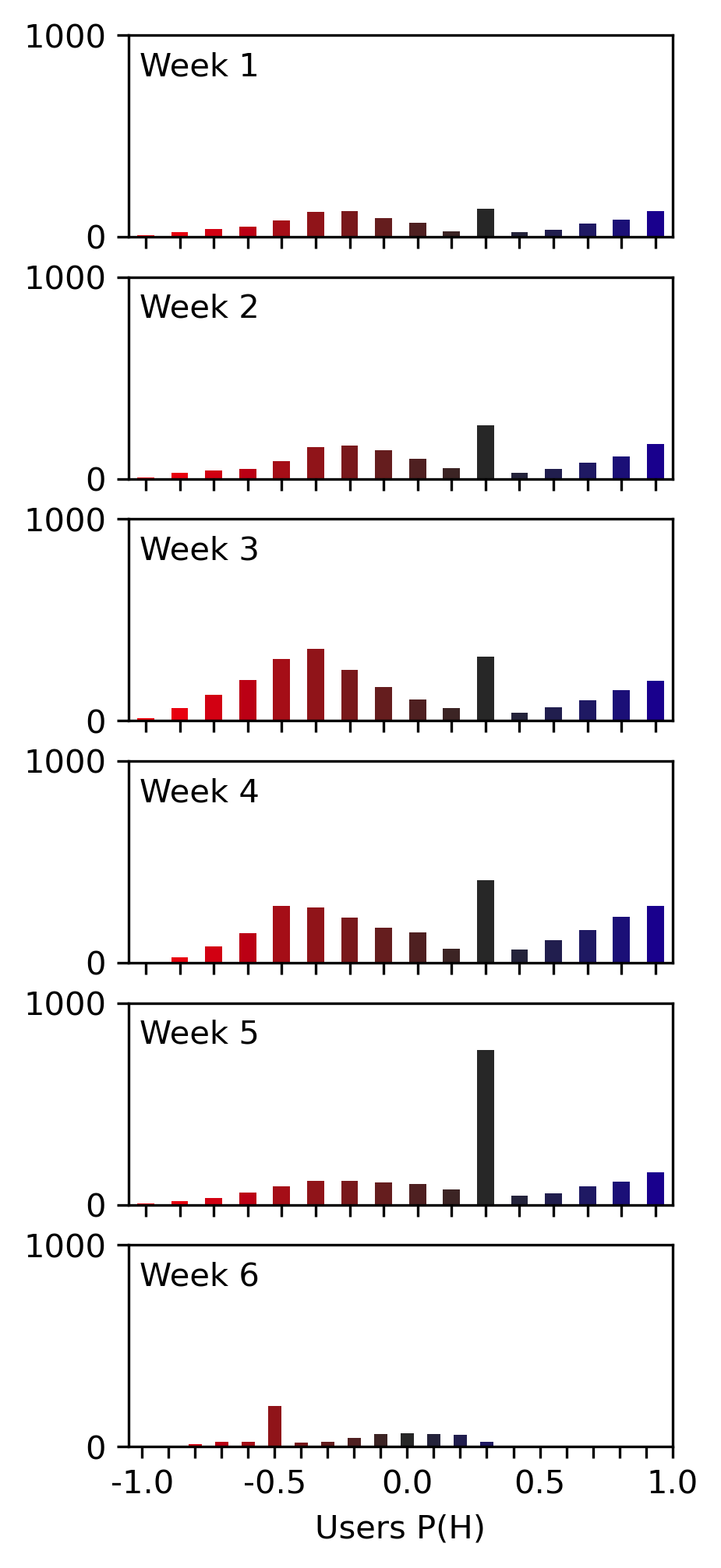"}
        \caption{Canada}
        \label{fig:polarity-can}
    \end{subfigure}
    \caption{Number of users by polarity histograms.}
    \label{fig:polarity}
\end{figure}

To provide general context, we first analyze the global polarization of each dataset; the metrics of variance, kurtosis, and polarization degree were extracted based on the $P(H)$ value of the users who retweeted each week; the analysis is inspired by \cite{dimaggio1996have}. Variance refers to the dispersion of opinions among individuals. The more dispersed the opinions become, the more difficult it becomes for the political system to maintain consensus \cite{dimaggio1996have}. Kurtosis refers to the bimodality in the distribution of opinions among individuals. Positive values mean consensus, and negative values, as it gets closer to $-2.0$, mean disagreement among individuals. The more these curves diverge, the greater the probability of social disagreement \cite{dimaggio1996have}. The polarization degree, calculated by the average absolute values of $P(H)$ for all users, refers to how much the opinions of individuals tend to one or both extremes in the $P(H)$ spectrum: a value around $0.0$ means that most users tend to be ideologically neutral, while a value around $1.0$ means a more extreme bias to one or both sides of the political spectrum. Table~\ref{tab:polarity} shows the values for these metrics, and Figure~\ref{fig:polarity} shows the weekly histograms of users by $P(H)$ range, divided into 21 bins.

\begin{table}[ht!]
    \centering
    \caption{\textbf{Polarization analysis.} P(H) summarized variables: variance (var), kurtosis (kurt) and polarization degree (deg).}
    \label{tab:polarity}
    \begin{tabular}{c|rrr|rrr}
        \hline
        Week & \multicolumn{3}{c|}{Brazil} & \multicolumn{3}{c}{Canada} \\
        
             & Var   & Kurt   & Deg   & Var   & Kurt   & Deg   \\
        \hline
   1 &      0.439 &    -1.494 &   0.621 &      0.298 &    -0.890 &   0.593 \\
   2 &      0.481 &    -1.438 &   0.659 &      0.280 &    -1.059 &   0.543 \\
   3 &      0.597 &    -1.774 &   0.743 &      0.321 &    -1.514 &   0.511 \\
   4 &      0.507 &    -0.993 &   0.775 &      0.283 &    -1.341 &   0.492 \\
   5 &      0.155 &    -0.473 &   0.387 &      0.220 &    -0.967 &   0.393 \\
   6 &      0.279 &    -0.706 &   0.594 &      0.099 &    -1.063 &   0.310 \\
        \hline
    \end{tabular}
\end{table}

\begin{figure}[ht!]
    \centering
    \begin{subfigure}{6cm}
        \centering
        \includegraphics[width=6cm]{"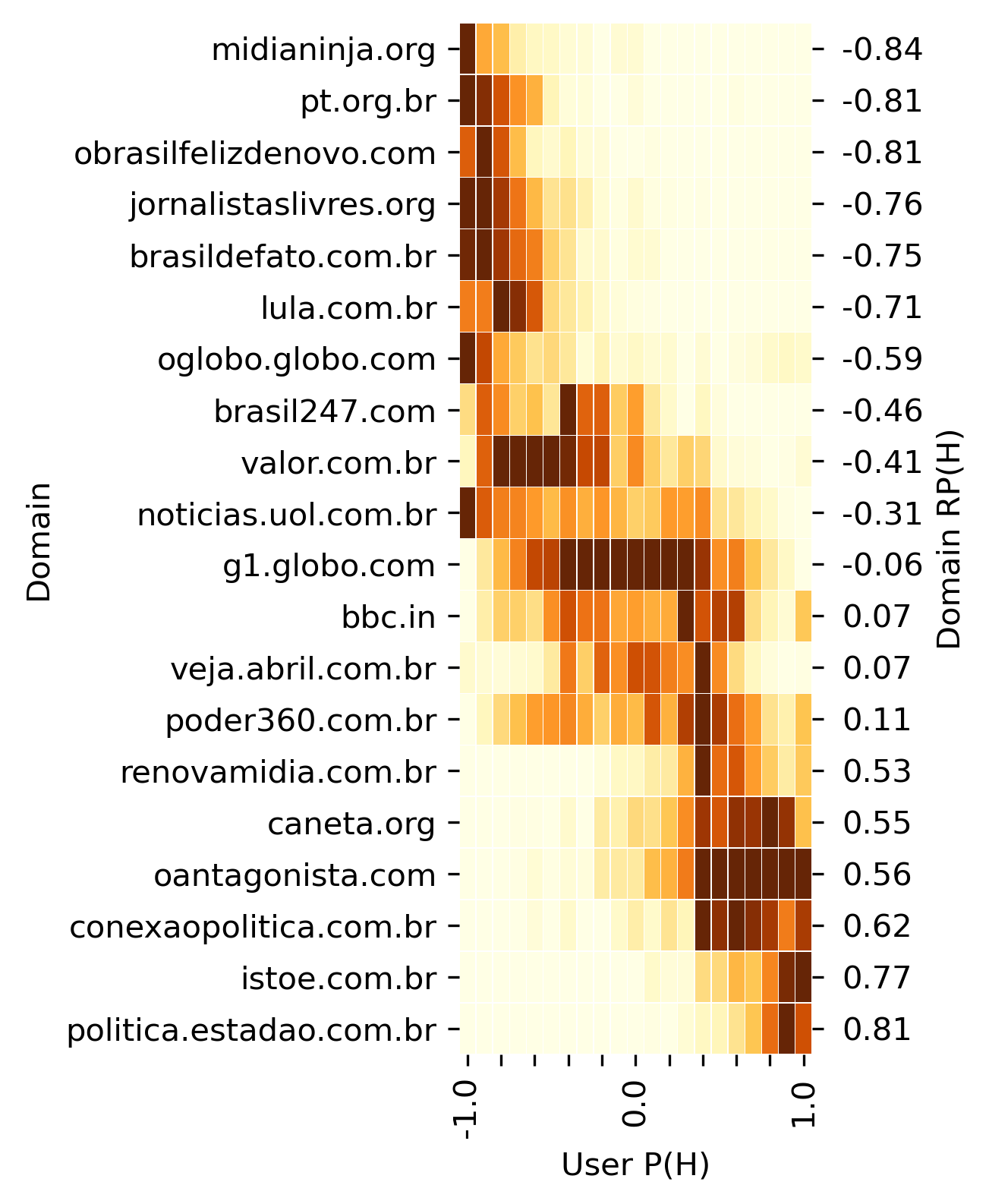"}
        \caption{Brazil}
        \label{fig:domain_heatmaps-bra}
    \end{subfigure}%
    \begin{subfigure}{6cm}
        \centering
        \includegraphics[width=5.7cm]{"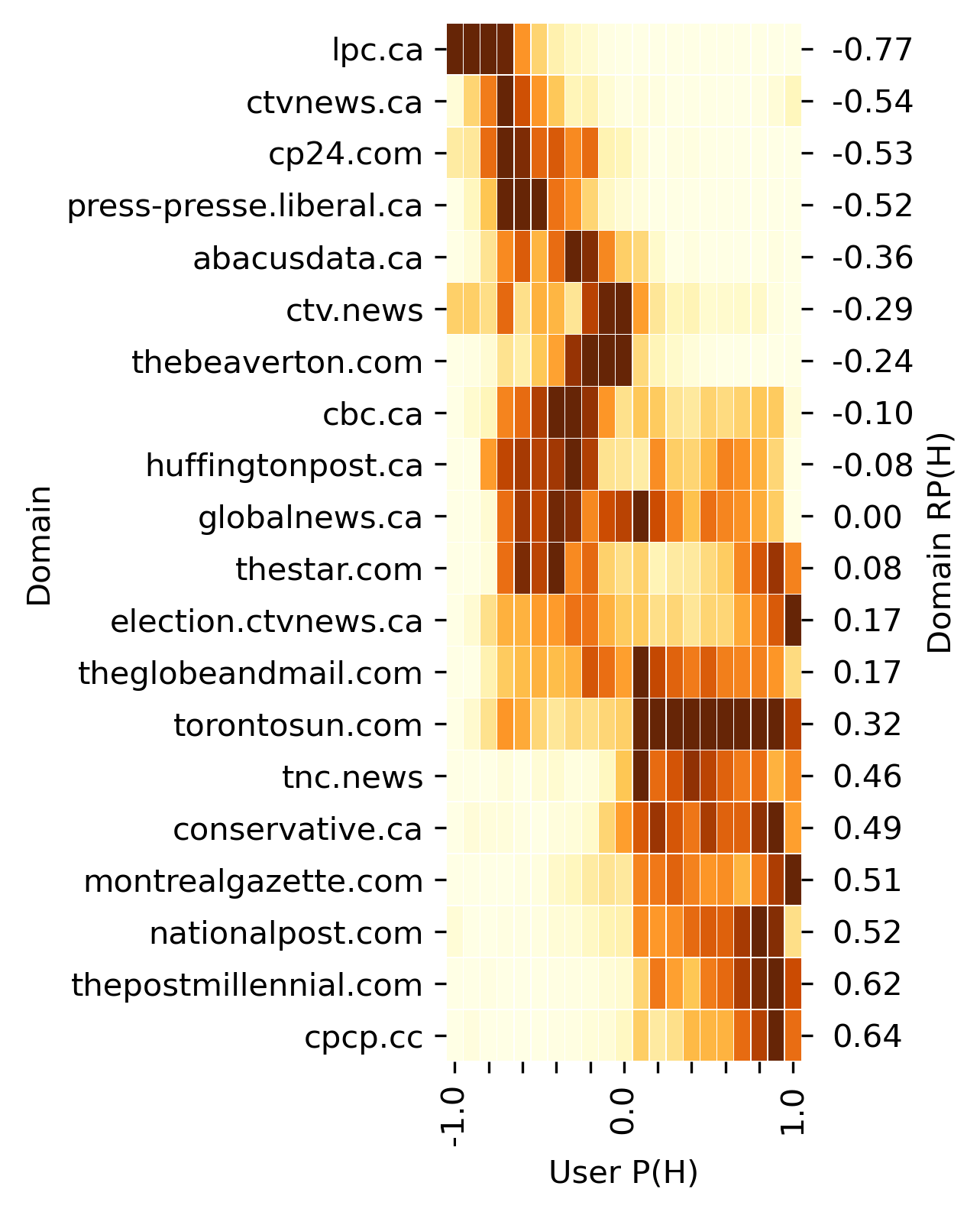"}
        \caption{Canada}
        \label{fig:domain_heatmaps-can}
    \end{subfigure}
    \caption{User engagement with domains.}
    \label{fig:domain_heatmaps}
\end{figure}

In both datasets, it is possible to visually identify bimodal curves in the first four weeks. These weeks represent the period before the elections. The presence of the bimodal curve indicates that both datasets present evidence of polarization \cite{Fiorina2008, dimaggio1996have}, with the polarization in the Brazilian case stronger than in the Canadian case. The variance indicates that the Brazilian case had a greater dispersion in political opinions, while it was less dispersed in Canada. Right-wing radicals in both situations appear proportionally more concentrated than left-wing radicals, which could be interpreted as the former being more prone to retweet content shared by the top \textit{bubble reachers} than the latter. Finally, the polarization degree indicates a more extreme scenario in Brazil, mainly in week 4, which refers to the week before the voting day. After the election in Brazil, polarization became lower on both left and right-wing sides. The histograms show some polarization in Canada, but with a significant amount of neutral users during the entire observed period. However, weeks 5 and 6 exhibit proportionally more neutral users than polarized users than previous weeks, and a substantial reduction of polarized users occurs after the election result (week 6). 

To provide validation of these polarization metrics on the network, we conduced a network assortativity analysis for all retweet networks representing each week. Table~\ref{tab:assortativity} summarize the assortativity regarding the similarity of connections in the networks considering the node orientation ($L$, $N$, or $R$) for Brazil and Canada. The first column "All nodes" considers the entire network, where it is possible to observe a higher polarization in Brazil than in Canada. However, when we remove $N$ labeled nodes – second column "L and R nodes" – both countries show a very high assortativity. As $N$ nodes tend to act as bridges in several situations – especially in the Canada dataset – the polarization is more evident when we remove them.

\begin{table}[ht!]
    \caption{Assortativity on retweet networks according to node's political orientation.}
    \label{tab:assortativity}
    \centering
    \begin{tabular}[h]{c|cc|cc}
    \hline
    & \multicolumn{2}{c}{Brazil} & \multicolumn{2}{c}{Canada} \\
    \cmidrule(lr){2-3} \cmidrule(lr){4-5}
    Week & All nodes & L and R nodes & All nodes & L and R nodes \\
    \hline
    1 & 0.68 & 0.96 & 0.48 & 0.94 \\
    2 & 0.74 & 0.94 & 0.49 & 0.98 \\
    3 & 0.82 & 0.96 & 0.48 & 0.97 \\
    4 & 0.85 & 0.97 & 0.47 & 0.95 \\
    5 & 0.31 & 0.78 & 0.48 & 0.98 \\
    6 & 0.48 & 0.80 & 0.50 & 0.27 \\
    \hline
    \end{tabular}
\end{table}

This high degree of overall polarization, even among users exposed to the content shared by the top bubble reachers, is an initial sign that exposure to cross-cutting content alone is not enough to remedy polarization, especially in the emotionally intensified context of a national election. This finding speaks to our second research question.

\subsection{Most central users characterization}

\begin{figure}[ht!]
    \centering
    \begin{subfigure}{6cm}
        \centering
        \includegraphics[width=6cm]{"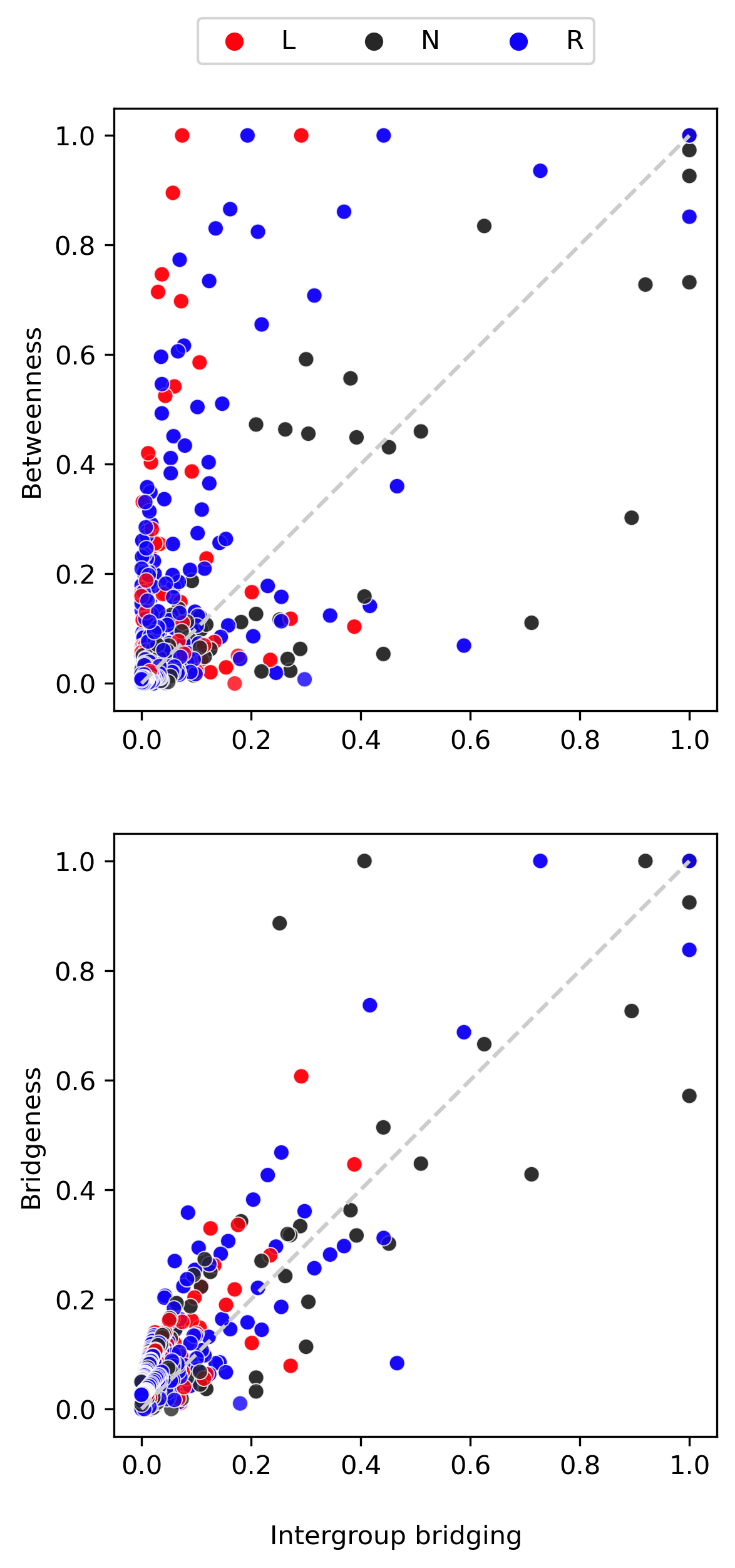"}
        \caption{Brazil}
        \label{fig:top_100_centralities_comparison-bra}
    \end{subfigure}%
    \begin{subfigure}{6cm}
        \centering
        \includegraphics[width=6cm]{"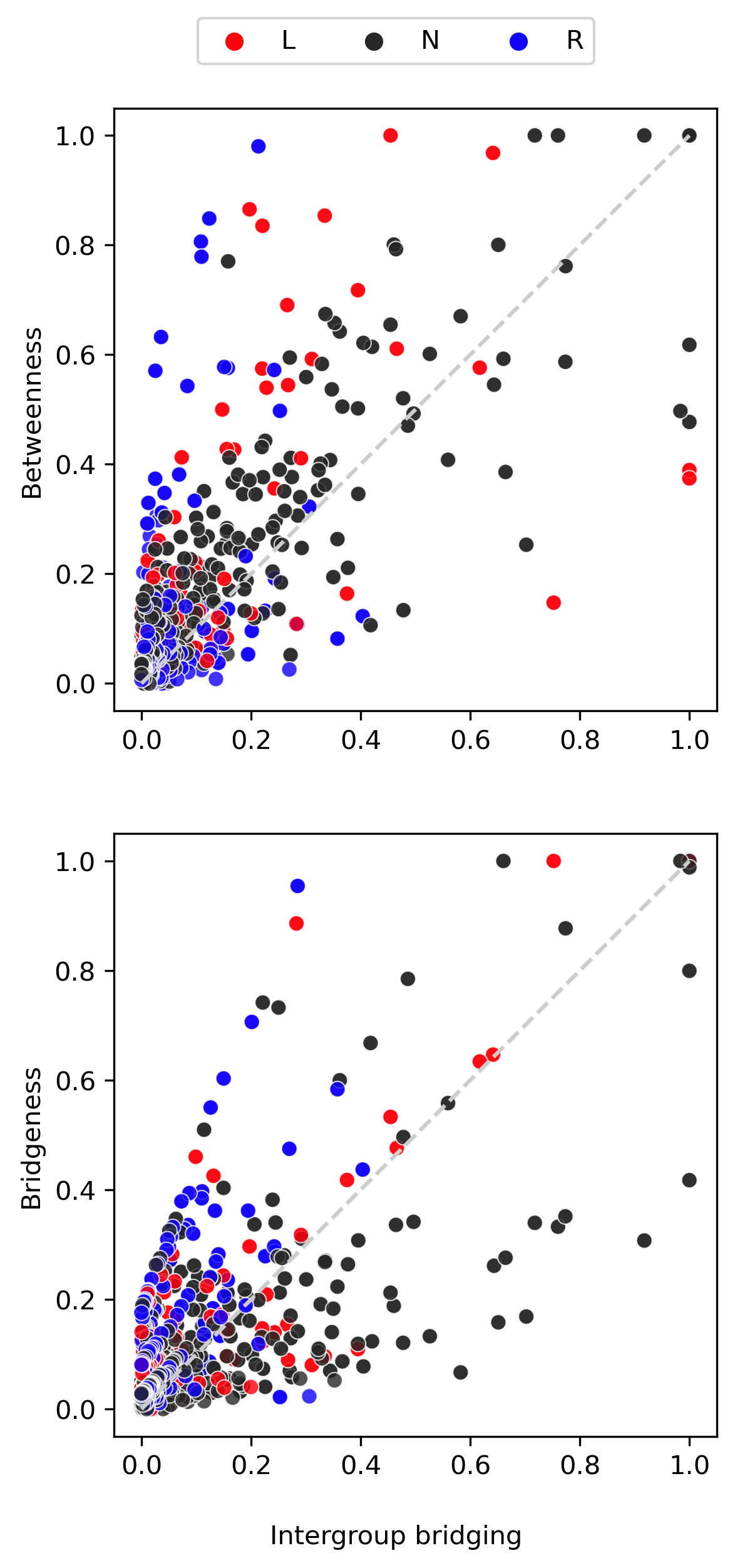"}
        \caption{Canada}
        \label{fig:top_100_centralities_comparison-can}
    \end{subfigure}
    \caption{Comparison between centrality metrics in all weeks, with top 100 users according to \textit{intergroup bridging} and the top 100 users according to betweenness or bridgeness.}
    \label{fig:top_100_centralities_comparison}
\end{figure}

In order to highlight the difference of the \textit{intergroup bridging} metric compared to betweenness and bridgeness, the relationships between these centralities were compared in Figure~\ref{fig:top_100_centralities_comparison}. Scatter plots express the relationship between \textit{intergroup bridging} and the other metrics, where each point represents accounts among the top $100$ users according to \textit{intergroup bridging} and the top 100 users according to betweenness or bridgeness centralities. The X-axis is the value of the \textit{intergroup bridging} centrality, and the Y-axis is the value of the betweenness or bridgeness centrality for each data point – values are normalized on the $[0.0;1.0]$ interval for each metric in order to allow a fair comparison between metrics. Colors indicate user's political orientation, being red for L, black for N, and blue for R. By looking at the scatter plots, it is evident that \textit{intergroup bridging} does not classify as the most central the same nodes as betweenness and bridgeness. It is possible to observe a tendency for \textit{intergroup bridging} to give higher values to N-labeled nodes, disfavoring L and R. \textcolor{black}{It's worth to mention that neutral users, in this case, are less averse to connect to out-group members than partisans by retweeting or being retweeted, thus, not impacting these patterns observed for \textit{intergroup bridging}.} Another important fact – not visible in the figures – is that \textit{intergroup bridging} tends to rank higher verified accounts compared to bridgeness - a critical scenario - among the top $100$ accounts in all weeks, \textit{intergroup bridging} identified $77$ verified ones against $55$ by bridgeness in Brazil. Where in Canada, \textit{intergroup bridging} identified $152$ versus $110$ by bridgeness. Being able to rank higher verified accounts is interesting because the algorithm better capture accounts less probable to be bots, which tends to be central in the diffusion of information \cite{gonzalez2021bots}.

\begin{figure}[ht!]
    \centering
    \begin{subfigure}{6cm}
        \centering
        \includegraphics[width=6cm]{"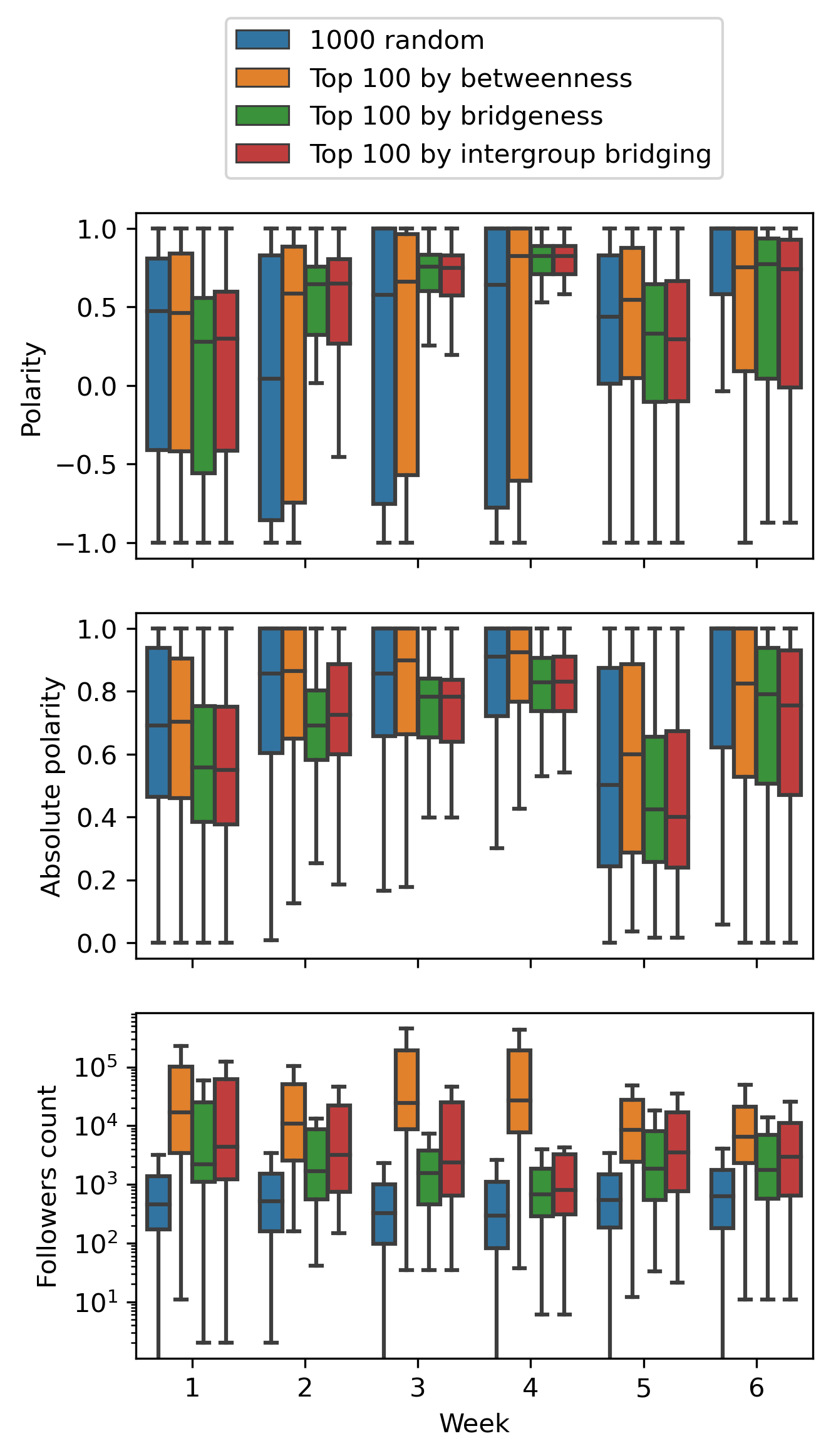"}
        \caption{Brazil}
        \label{fig:top_bubble_poppers_attributes-bra}
    \end{subfigure}%
    \begin{subfigure}{6cm}
        \centering
        \includegraphics[width=6cm]{"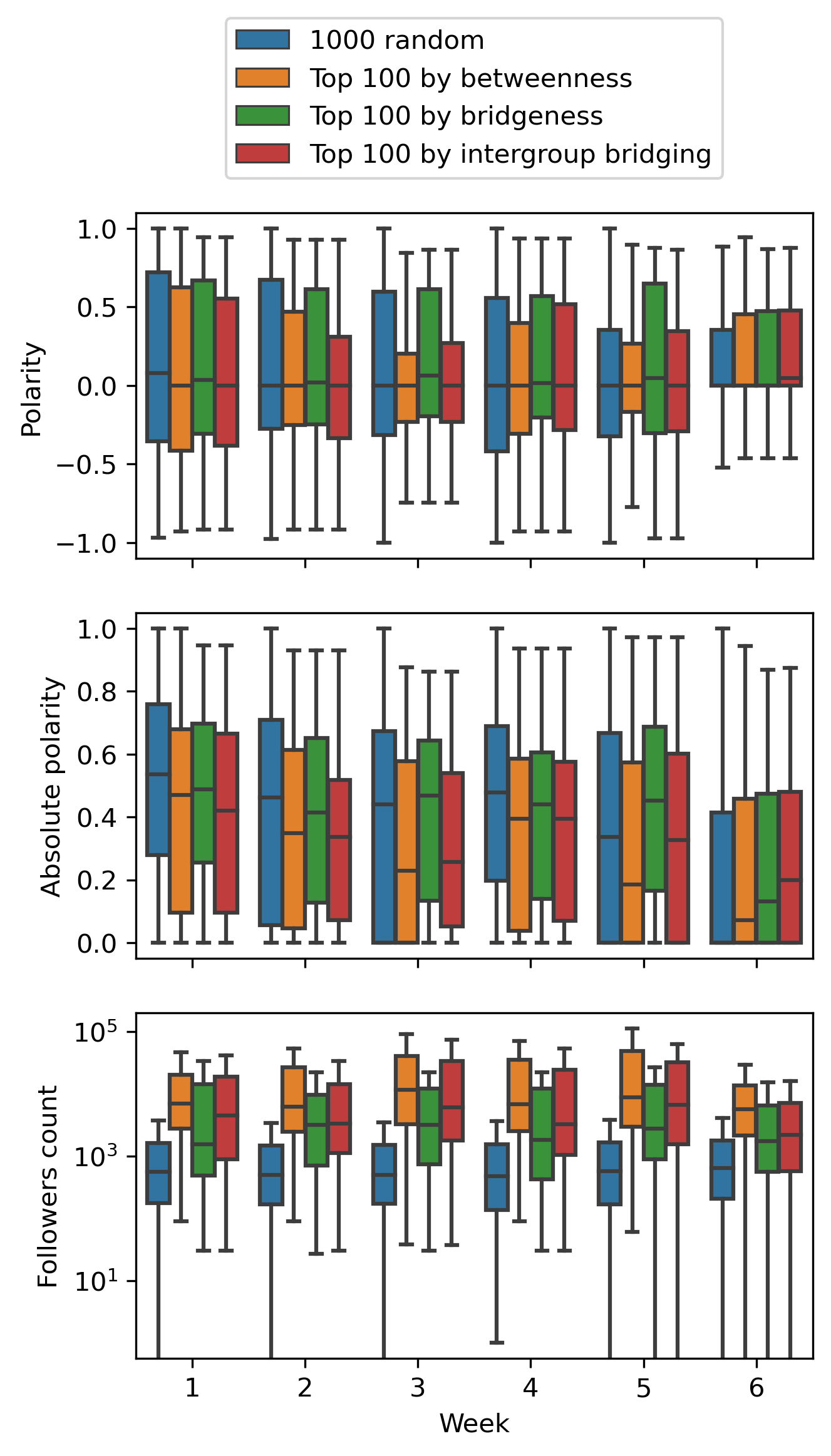"}
        \caption{Canada}
        \label{fig:top_bubble_poppers_attributes-can}
    \end{subfigure}
    \caption{Characterization of top 100 users by each centrality metric.}
    \label{fig:top_bubble_poppers_attributes}
\end{figure}

To provide descriptive information about the top $100$ user accounts by each centrality, we explored the following metrics: (i) ``polarity,'' measured by the users' average polarity, thus, representing the ideological distribution of users; (ii) ``polarization degree,'' measured by the average of the absolute values of polarity, representing the degree of extremism in direction to one or both sides in the ideological distribution; and (iii) ``followers count,'' indicating the degree of influence of these accounts through the number of followers they have. To compare these metrics, four groups of users were selected in each of the weeks: top $100$ users by the betweenness metric, top $100$ users by the bridgeness metric, top $100$ users by the \textit{intergroup bridging} metric, and $1000$ random users. This information was compiled in Figure~\ref{fig:top_bubble_poppers_attributes}. Each graph represents one of the metrics, the horizontal axis the weeks in the dataset, and the vertical axis the value of the respective metric on each graph. Analyzing this figure, it is clear that all metrics rank certain users differently. For simplicity, we will concentrate the analysis between \textit{intergroup bridging} and bridgeness. In our dataset, \textit{intergroup bridging} ranked higher users with average polarity lower than bridgenness. This is also the case when we look at the absolute polarity metric, where top users by \textit{intergroup bridging} are less ideologically extremist. More central users according to \textit{intergroup bridging} tend to have more followers in both datasets, indicating that it identify, in general, more influential users.

\begin{figure}[ht!]
    \centering
    \begin{subfigure}{6cm}
        \centering
        \includegraphics[width=6cm]{"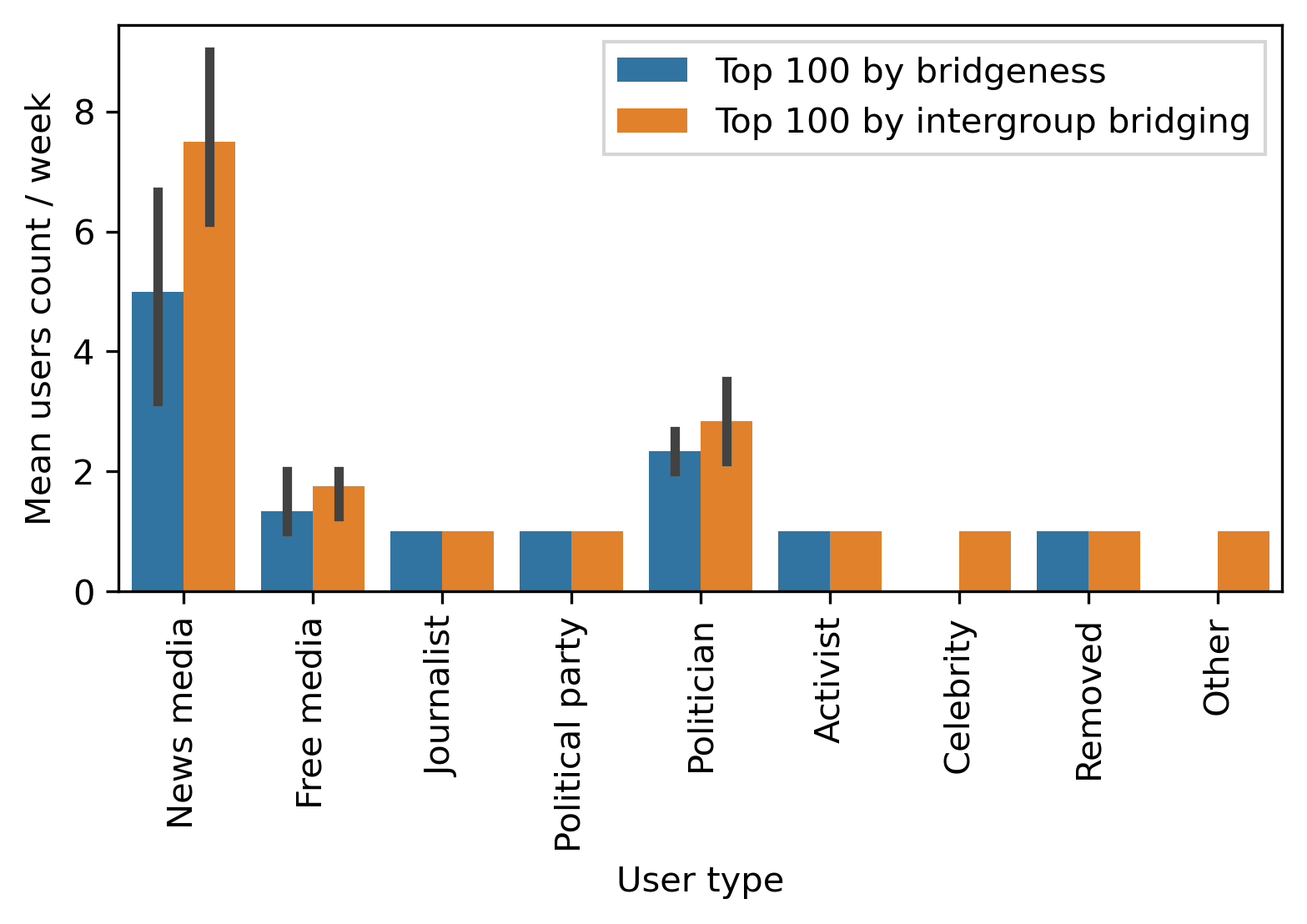"}
        \caption{Brazil}
        \label{fig:verified_accounts_type-bra}
    \end{subfigure}%
    \begin{subfigure}{6cm}
        \centering
        \includegraphics[width=5.7cm]{"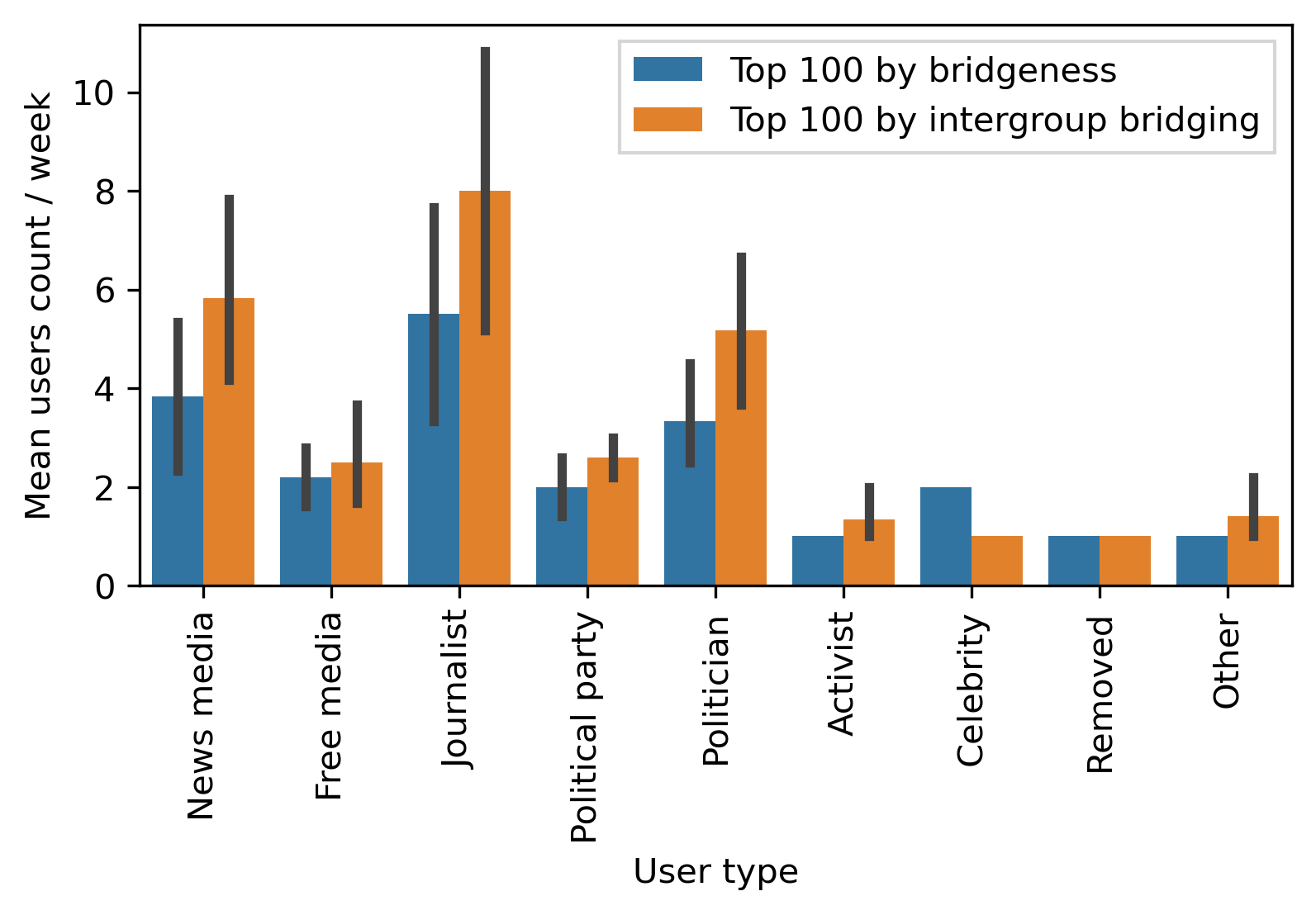"}
        \caption{Canada}
        \label{fig:verified_accounts_type-can}
    \end{subfigure}
    \caption{Mean number of accounts per week according to its type. This analysis considers only verified accounts among the top 100 verified ones in all weeks by each centrality metric.}
    \label{fig:verified_accounts_type}
\end{figure}

Considering the importance of the most central users for this study, they were manually labeled according to its type to understand on what proportion they appear on the dataset. User type could be: ``news media'', for communication channels that focus on the transmission of facts, seeking to avoid positioning bias, ``free media'', for communication channels that allow the expression of opinions freely and spontaneously, ``journalist'', for journalism professionals, ``political party'', for political organizations or users representing a political party, ``politician'', for candidates, individuals with political office or ex-politicians, `` activist'', for politically motivated users who defend a political position, party or candidate, ``celebrity'', for individuals with notable public knowledge, ``removed'', for accounts removed from Twitter, and ``other'', for all other cases not covered by any other type. While it was not feasible to manually label all top $100$ users found in all weeks in both countries (which was around $900$ unique accounts), this analysis was concentrated on verified accounts from that set, being $61$ for Brazil and $87$ for Canada). Figure~\ref{fig:verified_accounts_type} shows the mean number of accounts per week according to their type. It is possible to note that \textit{intergroup bridging} finds more verified accounts than bridgeness. Most of these accounts are news media or journalists (especially in Canada – most of them are connected to a news media). These results also reinforce that the novelties introduced by \textit{intergroup bridging} help to rank different users. Also, the fact that \textit{intergroup bridging} and those accounts identify the majority of verified accounts tend to be connected to news media, one would expect that they are essential for the phenomenon of ideology bridge that we envision to study, suggesting an advantage of this metric.

\begin{figure}[ht!]
    \centering
    \begin{subfigure}{6cm}
        \centering
        \includegraphics[width=6cm]{"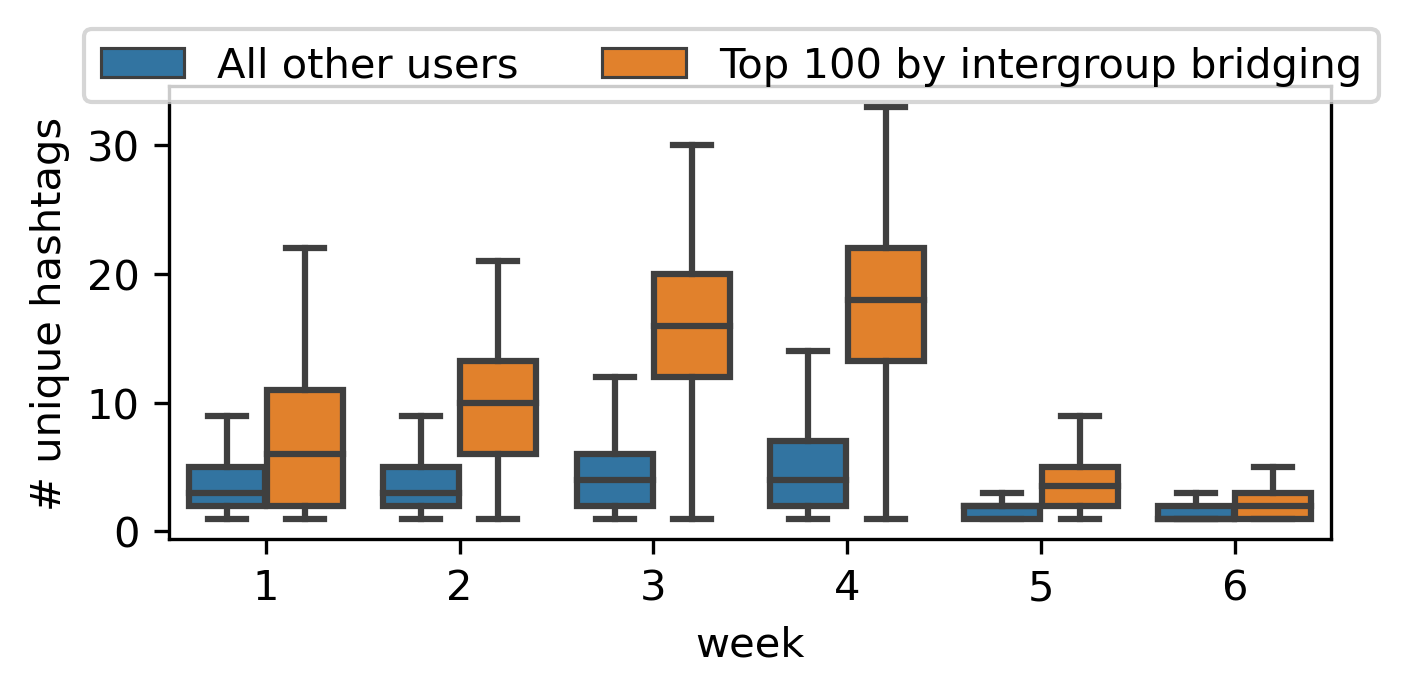"}
        \caption{Brazil}
        \label{fig:hastags_count_by_user-bra}
    \end{subfigure}%
    \begin{subfigure}{6cm}
        \centering
        \includegraphics[width=5.7cm]{"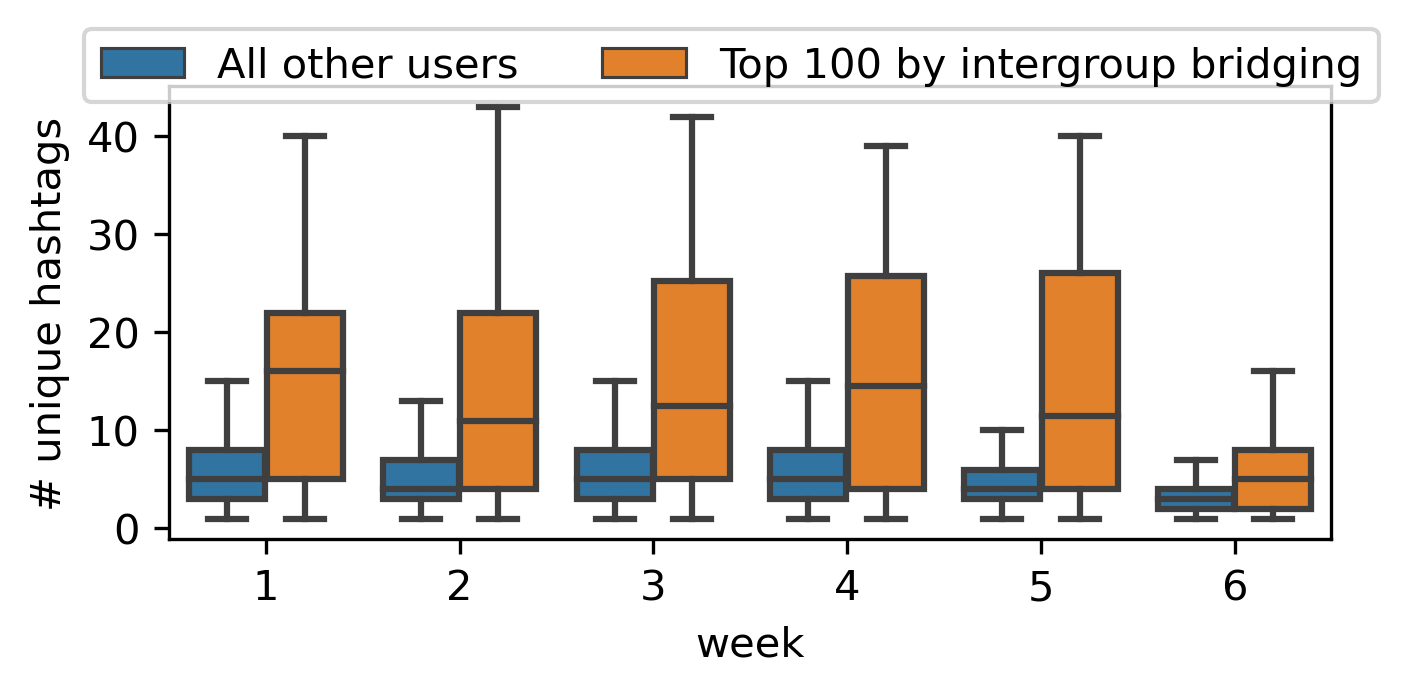"}
        \caption{Canada}
        \label{fig:hastags_count_by_user-can}
    \end{subfigure}
    \caption{Mean unique hashtags count by user.}
    \label{fig:hastags_count_by_user}
\end{figure}

The diversity of content shared by highly central users according to \textit{intergroup bridging} centrality was also analyzed. For this, the number of unique hashtags was used as an indicator of diversity. We found that top 100 \textit{bubble reachers} tend to use more unique hashtags than all other users – see Figure~\ref{fig:hastags_count_by_user}. This suggests that central \textit{bubble reachers} share more diverse content. \textcolor{black}{It is worth mentioning that the number of unique hashtags applied according to its political orientation is still virtually the same between highly central bubble reachers and all other users. That is, they do not have clear differences in choosing hashtags according to its political orientation.}

\subsection{User engagement}

This section discusses how users with specific political orientations engage by retweeting messages that support their political biases in different perspectives: domain, content, and topic. For this, we examine the correlation between the user P(H) who made a retweet and the RP(H) of the domain, content, and topics that were retweeted. The results of the correlations are shown in Table~\ref{tab:correlations_1}, where columns refer to the variables that were tested, with ``User P(H)'' being the P(H) value for the user who made a retweet, and ``Domain RP(H)'', ``Content RP(H)'', and ``Topic RP(H)'' the RP(H) value of the retweeted entity. To structure the tests, we considered that users classified with a neutral political orientation (with the label ``$N$'') behave differently from polarized users (with labels ``$L$'' or ``$R$''). For this reason, three tests were created, each with a distinct set of retweets filtered by the user P(H): (1) including all retweets, regardless of the user’s political orientation, (2) including only retweets made by neutral users and (3) including only retweets made by polarized users. See Table~\ref{tab:correlations_1}.

Considering that an entity's RP(H) value is defined based on the P(H) of the users who retweeted it, we evaluated the possible existence and impact of circularity in the correlation tests. For this purpose, a synthetic dataset of random retweets was created for each performed correlation test, maintaining the same number of retweets, users, and entities present in the original retweets dataset, but with P(H) values generated randomly following a uniform distribution. Each synthetic dataset was generated 100 times with different random seeds. For each synthetic dataset, we generate two null models: Model A with RP(H) being calculated following the method presented in section \ref{tab:correlations_1} and Model B with RP(H) generated randomly following a uniform distribution. As expected for a full random model, Model B showed no correlation between the P(H) of users and the RP(H) of domains, content, or topics for all tests. On the other hand, Model A showed weak signs of positive correlations, so we focused on presenting results from this model.

\begin{table}[ht!]
  \caption{Correlation between User P(H) and retweeted Domain, Content or Topic RP(H).}
  
  \label{tab:correlations_1}
  \resizebox{\columnwidth}{!}{%
  \begin{tabular}[h]{l|lll|lll}
               & \multicolumn{3}{c|}{Brazil}                 & \multicolumn{3}{c}{Canada}                 \\
               \cmidrule(lr){2-4} \cmidrule(lr){5-7}
     User P(H) & Domain RP(H) & Content RP(H) & Topic RP(H) & Domain RP(H) & Content RP(H) & Topic RP(H) \\
    \hline
    (1) all &  \makecell[l]{$N = 23,033$\\$r = 0.834$***\\$\rho = 0.790$***} &  \makecell[l]{$N = 23,844$\\$r = 0.892$***\\$\rho = 0.833$***} &  \makecell[l]{$N = 18,484$\\$r = 0.643$***\\$\rho = 0.619$***} &  \makecell[l]{$N = 31,070$\\$r = 0.701$***\\$\rho = 0.601$***} &  \makecell[l]{$N = 31,370$\\$r = 0.878$***\\$\rho = 0.721$***} &  \makecell[l]{$N = 21,487$\\$r = 0.381$***\\$\rho = 0.332$***} \\
    \hline
    (2) neutral &   \makecell[l]{$N = 2,960$\\$r = 0.404$***\\$\rho = 0.385$***} &   \makecell[l]{$N = 2,956$\\$r = 0.512$***\\$\rho = 0.507$***} &   \makecell[l]{$N = 2,445$\\$r = 0.359$***\\$\rho = 0.360$***} &   \makecell[l]{$N = 6,278$\\$r = 0.448$***\\$\rho = 0.410$***} &   \makecell[l]{$N = 6,261$\\$r = 0.625$***\\$\rho = 0.599$***} &   \makecell[l]{$N = 4,420$\\$r = 0.216$***\\$\rho = 0.220$***} \\
    \hline
    (3) polarized &  \makecell[l]{$N = 20,073$\\$r = 0.857$***\\$\rho = 0.783$***} &  \makecell[l]{$N = 20,888$\\$r = 0.913$***\\$\rho = 0.816$***} &  \makecell[l]{$N = 16,039$\\$r = 0.666$***\\$\rho = 0.625$***} &  \makecell[l]{$N = 24,792$\\$r = 0.727$***\\$\rho = 0.518$***} &  \makecell[l]{$N = 25,109$\\$r = 0.913$***\\$\rho = 0.619$***} &  \makecell[l]{$N = 17,067$\\$r = 0.396$***\\$\rho = 0.293$***} \\
\hline
\end{tabular}%
}
    \caption*{Note: $r$ and $\rho$ represent the Pearson and Spearman's correlation coefficients, respectively. All correlations are significant at $p<.001$ level (two-tailed).}
\end{table}

\textbf{User P(H) vs domain RP(H):} Considering all retweets, a strong to very strong positive correlation was found, with $r(23,033) = .834$***, and $\rho(23,033) = .790$*** in the Brazilian dataset and a strong positive correlation with $r(31,070) = .701$***, and $\rho(31,070) = .601$***, in the Canadian dataset. When looking only at neutral users, this correlation is slightly less intense than polarized users in both datasets. This result indicates that, in general, users tend to retweet content from domains with a political orientation aligned to their own, even if they do not always do so. Figure~\ref{fig:domain_heatmaps} show the retweet density made by each user polarity from the top $20$ most retweeted domains on each dataset and their respective RP(H) -- domains were sorted in an increasing order by their RP(H) value. In this figure, it is possible to note that a relatively small number of domains could be reaching the bubbles more than others, such as ``noticias.uol.com.br'' and ``globalnews.ca''. Considering that these domains could be producing content that appeals to users from diverging political orientations, we also have to verify if partisans interact in a balanced way with content that goes towards or against their own political bias (we present results regarding this point in Section~\ref{subsec:impact_of_source}). The correlations found for the null Model A were $ r(31,542) = .043 $*** $(\pm.0037)$, $ r(10,532) = .013 $ $(\pm.0088)$ and $ r(21,010) = .052 $*** $(\pm.0045)$ for tests 1, 2 and 3, respectively, in Canada and $ r(23,885) = .031 $*** $(\pm.0047)$, $ r(7,961) = .007 $ $(\pm.01)$ and $ r(15,924) = .038 $*** $(\pm.0059)$, respectively, in Brazil. Values are similar for Spearman's correlations. We observe that there is a very low circularity that is negligible. Also note that the original correlation values are much larger than the limits of the null model, thus, no significant impact of circularity in the results.

\begin{figure}[ht!]
    \centering
    \begin{subfigure}{6cm}
        \centering
        \includegraphics[width=5cm]{"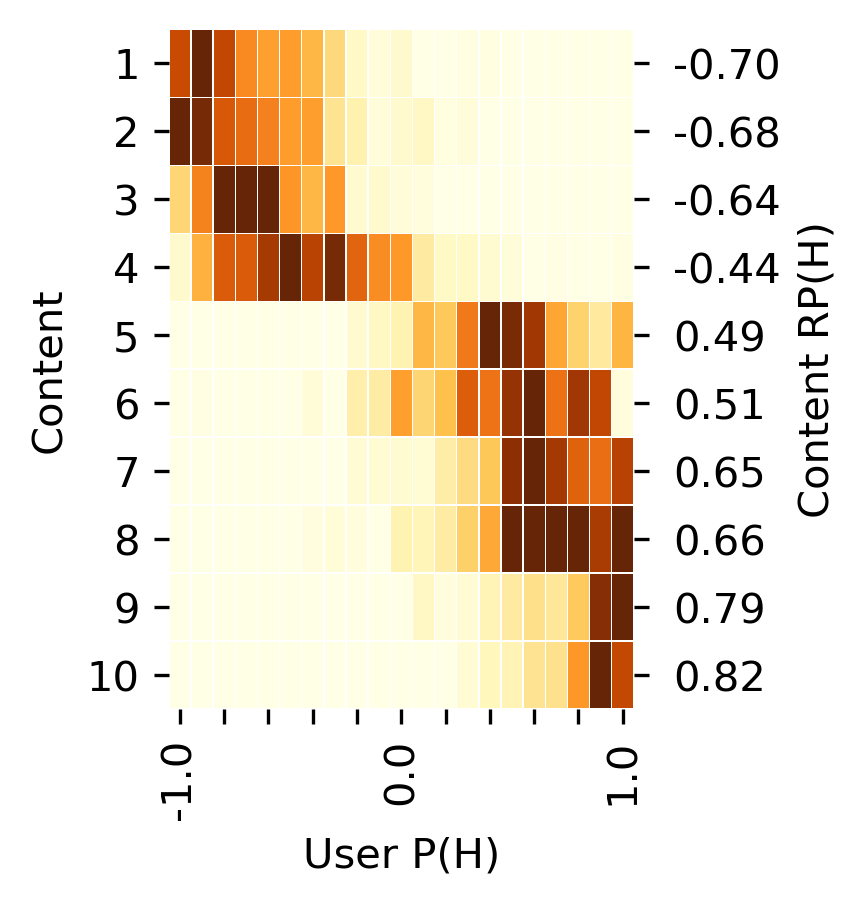"}
        \caption{Brazil}
        \label{fig:content_heatmap-bra}
    \end{subfigure}%
    \begin{subfigure}{6cm}
        \centering
        \includegraphics[width=5cm]{"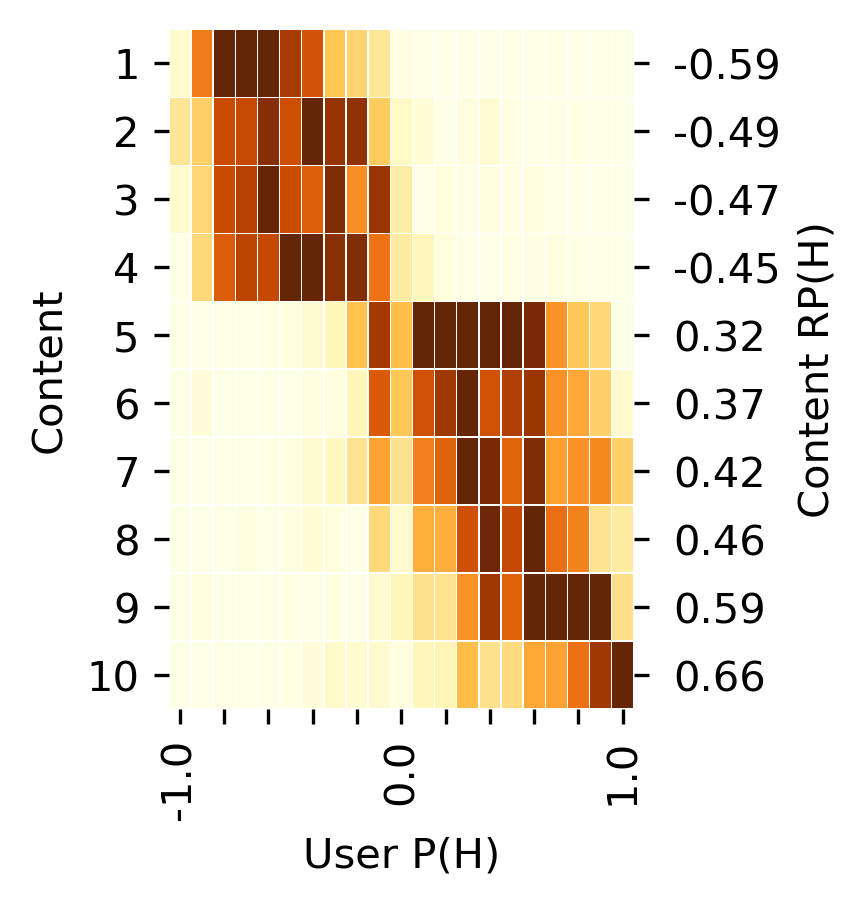"}
        \caption{Canada}
        \label{fig:content_heatmap-can}
    \end{subfigure}
    \caption{User engagement with news content.}
    \label{fig:content_heatmap}
\end{figure}

\begin{figure}[ht!]
    \centering
    \begin{subfigure}{6cm}
        \centering
        \includegraphics[width=5cm]{"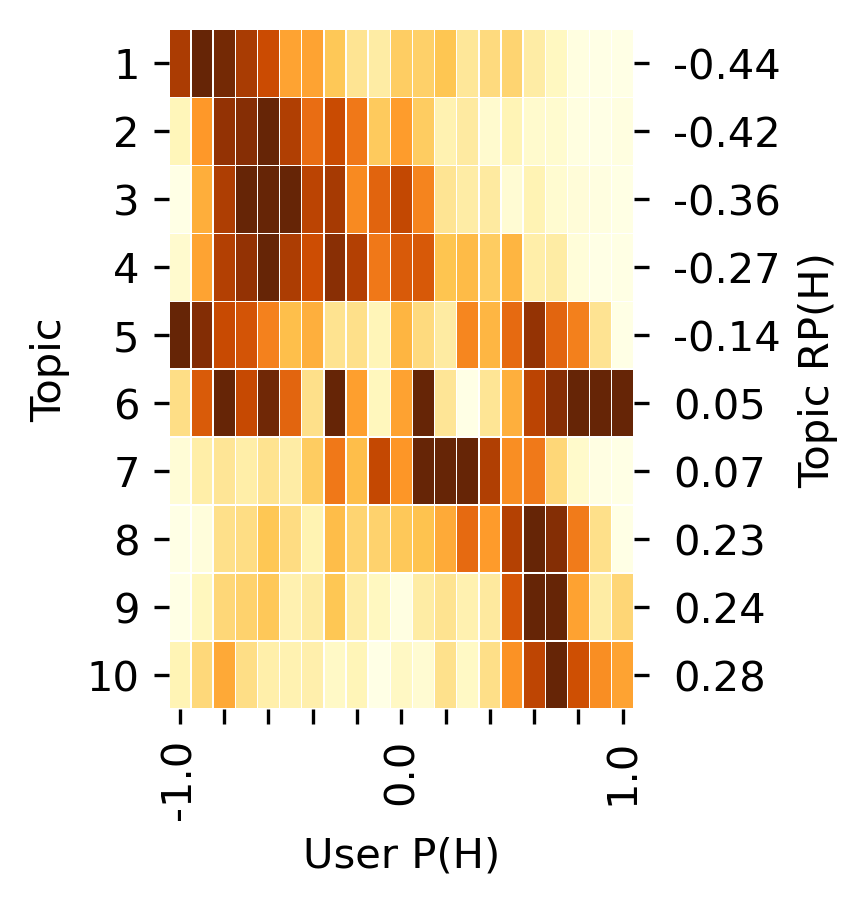"}
        \caption{Brazil}
        \label{fig:topic_heatmap-bra}
    \end{subfigure}%
    \begin{subfigure}{6cm}
        \centering
        \includegraphics[width=5cm]{"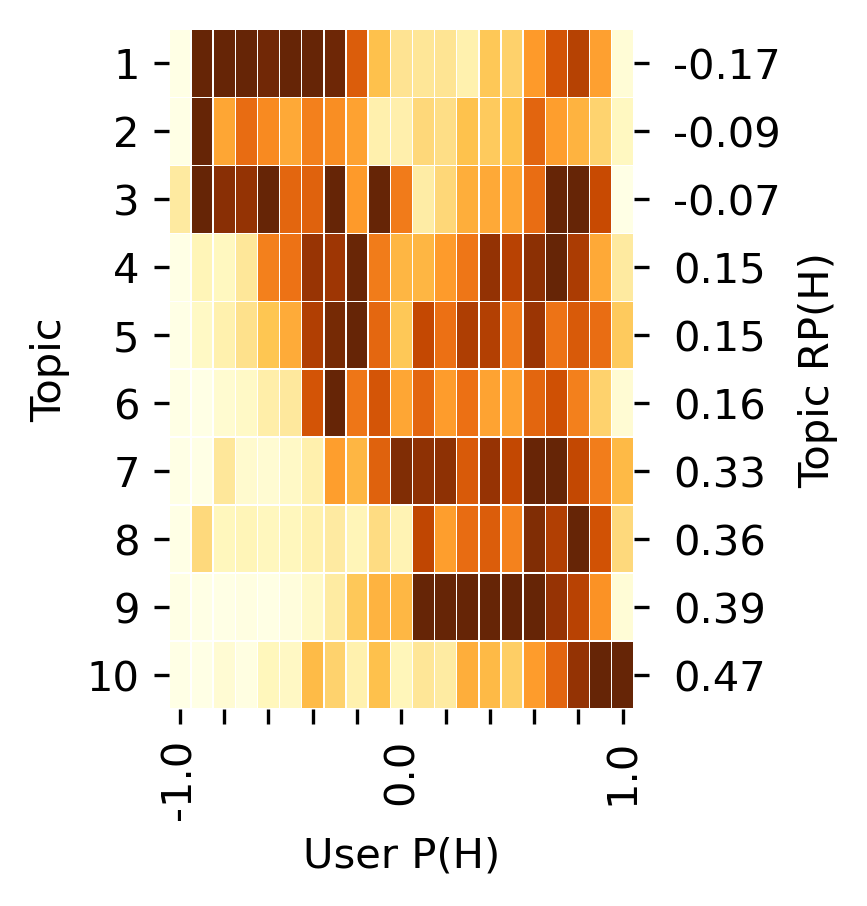"}
        \caption{Canada}
        \label{fig:topic_heatmap-can}
    \end{subfigure}
    \caption{User engagement with topics.}
    \label{fig:topic_heatmap}
    \begin{flushleft}
    \end{flushleft}
\end{figure}

\textbf{User P(H) vs content RP(H):} When considering all retweets, a strong to very strong positive correlation was identified in both datasets, with $r(23,844) = .892$***, and $\rho(23,844) = .833$*** for Brazil and $r(31,370) = .878$***, and $\rho(31,370) = .721$*** for Canada. The similarity between these results may indicate that, regardless of the dataset, most of the time, users engage with content that aligns with their political orientation. As with the correlation of user P(H) with domain RP(H), when neutral users are observed in isolation, the correlation is slightly less intense compared to polarized users. The fact that the correlation between user P(H) and content RP(H) is stronger than the correlation between user P(H) and domain RP(H) may indicate that the content’s political orientation is slightly more important than the domain political orientation in user retweet behaviour. Regardless, content shared by \textit{bubble reachers} remains highly polarized, and even ``neutral'' users engage with content in segmented ways. Figure~\ref{fig:content_heatmap} show the retweet density made by each user polarity from the top $10$ most retweeted URLs (content) on each dataset and their respective RP(H). URLs are ordered ascending according to their RP(H) value. In this figure, it is possible to note that none of the URLs reached the bubbles. Instead, the content was shared only by users with a political orientation that resonated with the content bias. This visualization indicates that, although some domains could be reaching the bubbles, users tend to select only the content that aligns with their beliefs, even when exposed to content that appeals to different groups. The correlations for the null Model A are $ r(31,542) = .150 $*** $(\pm.004)$, $ r(10,531) = .047 $*** $(\pm.01)$ and $ r(21,011) = .180 $*** $(\pm.0049)$ for tests 1, 2 and 3, respectively, in Canada and $ r(23,885) = .0.135 $*** $(\pm.0048)$, $ r(7,947) = .044 $*** $(\pm.013)$ and $ r(15,938) = .162 $*** $(\pm.0062)$, respectively, in Brazil. Values are similar to Spearman's correlations. Note that although the limits produced by this model are slightly higher than the previous case, the original correlation results are located well beyond the limits of observed circularity effects.

\textbf{User P(H) vs topics RP(H):} In the test for all retweets, a  weak to moderate positive correlation was found in both datasets, with $r(18,917) = .548$***, and $\rho(18,917) = .513$*** in the Brazilian dataset and $r(21,308) = .451$***, and $\rho(21,308) = .389$*** in the Canadian one. The similarity between the results may indicate that, regardless of the situation, the topic's political orientation has less influence on retweeting behaviour compared to the content or domain. This means that the content itself and where it appears is generally more important than what it says in terms of its topic. Figure~\ref{fig:topic_heatmap} shows the retweet density made by each user polarity from the top $10$ most retweeted topics on each dataset and its respective RP(H). Topics are sorted in an increasing order based on RP(H) value. It is possible to observe more topics slightly biased to the left in Brazil, while in Canada most of the topics are dispersed among partisans, regardless of their political orientation. The correlations for the null Model A are $ r(21,923)=.032 $*** $(\pm.0048)$, $ r(7,318)=.009 $ $(\pm.0116)$ and $ r(14,605) = .039 $*** $(\pm.0056)$ for tests 1, 2 and 3, respectively, in Canada and $ r(19,078)=.047 $*** $(\pm.0049)$, $ r(6,378)=.015 $ $(\pm.0113)$ and $ r(12,700) = .057 $*** $(\pm.0063)$, respectively, in Brazil. Values also are similar to Spearman's correlations. Here we also observe that the original correlations are located very far from the observed circularity limits. Therefore, these results, like the others, are also not due to pure circularity.

About our third research question, we found that more polarized users show stronger partisan alignment. However, relatively neutral users still tend to share content from media outlets that align with their own partisan beliefs.

\begin{table}[ht!]
    \caption{Correlation between retweeted Domain RP(H) and Content or Topic RP(H).}
    \label{tab:correlations_2}
    \centering
    \begin{tabular}[h]{l|ll|ll}
    \hline
                   & \multicolumn{2}{c}{Brazil}                 & \multicolumn{2}{c}{Canada}                 \\
                   \cmidrule(lr){2-3} \cmidrule(lr){4-5}
      Domain RP(H) & Content RP(H) & Topic RP(H) & Content RP(H) & Topic RP(H) \\
\hline
(1) all &  \makecell[l]{$N = 23,788$\\$r = 0.889$***\\$\rho = 0.876$***} &  \makecell[l]{$N = 19,029$\\$r = 0.677$***\\$\rho = 0.647$***} &  \makecell[l]{$N = 31,512$\\$r = 0.683$***\\$\rho = 0.610$***} &   \makecell[l]{$N = 21,186$\\$r = 0.343$***\\$\rho = 0.303$***} \\
\midrule
(2) neutral &  \makecell[l]{$N = 15,534$\\$r = 0.749$***\\$\rho = 0.740$***} &   \makecell[l]{$N = 8,922$\\$r = 0.439$***\\$\rho = 0.418$***} &  \makecell[l]{$N = 24,562$\\$r = 0.509$***\\$\rho = 0.454$***} &   \makecell[l]{$N = 19,657$\\$r = 0.196$***\\$\rho = 0.188$***} \\
\midrule
(3) polarized &   \makecell[l]{$N = 8,254$\\$r = 0.934$***\\$\rho = 0.766$***} &  \makecell[l]{$N = 10,107$\\$r = 0.726$***\\$\rho = 0.619$***} &   \makecell[l]{$N = 6,950$\\$r = 0.674$***\\$\rho = 0.806$***} &  \makecell[l]{$N = 1,529$\\$r = -0.159$***\\$\rho = -0.350$***} \\
    \hline
    \end{tabular}
    \caption*{Note: $r$ and $\rho$ represent the Pearson and Spearman's correlation coefficients, respectively. All correlations are significant at $p<.001$ level (two-tailed).}

\end{table}

\subsection{The impact of the content source}\label{subsec:impact_of_source}

The analyses that were carried out in the previous section indicate a stronger correlation between users’ polarity and content polarity than between users’ polarity and domain’ polarity. However, it is still not possible to know whether users who engage with domains that produce content that interests both sides, that is, domains with a neutral RP(H), retweet more content that reinforces their political views or tend to balance their retweets by including content that goes in favour and against their political orientation. To better understand this, correlation tests were carried out involving the retweeted domain RP(H) and the retweeted content or topic RP(H) produced by the domain. The results are shown in Table ~\ref{tab:correlations_2}. The tests were conducted in the same way as the previous ones, separating the dataset according to the domain RP(H), including: (1) all retweets, regardless of the retweeted domain RP(H), (2) only retweets from domains with neutral RP(H) and (3) only retweets from domains with polarized RP(H).

\begin{figure}[ht!]
    \centering
    \begin{subfigure}{6cm}
        \centering
        \includegraphics[width=5cm]{"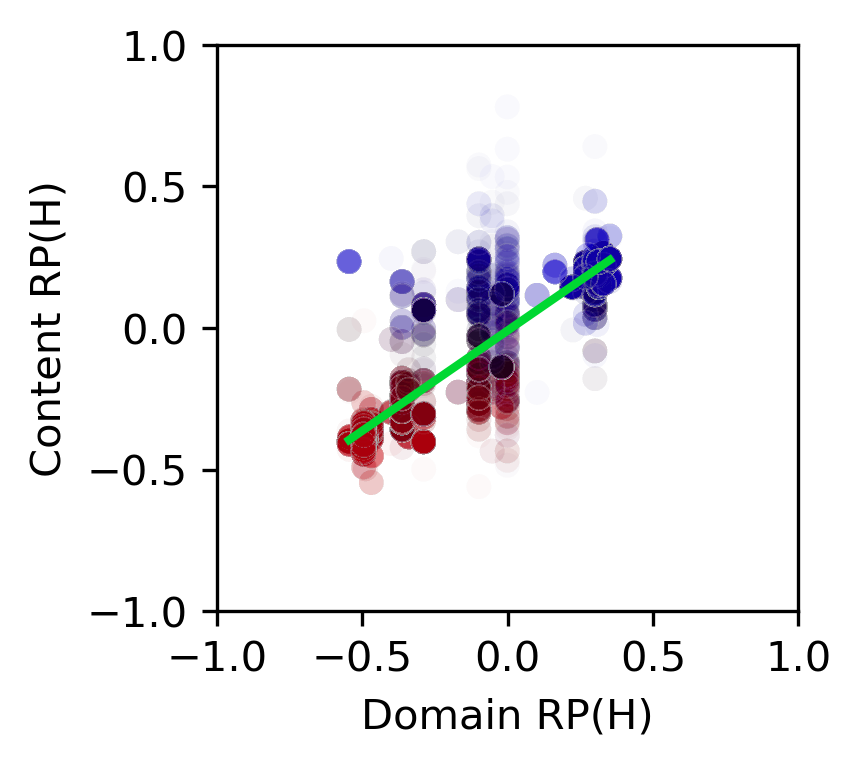"}
        \caption{Brazil}
        \label{fig:domain_vs_content_scatter-bra}
    \end{subfigure}%
    \begin{subfigure}{6cm}
        \centering
        \includegraphics[width=5cm]{"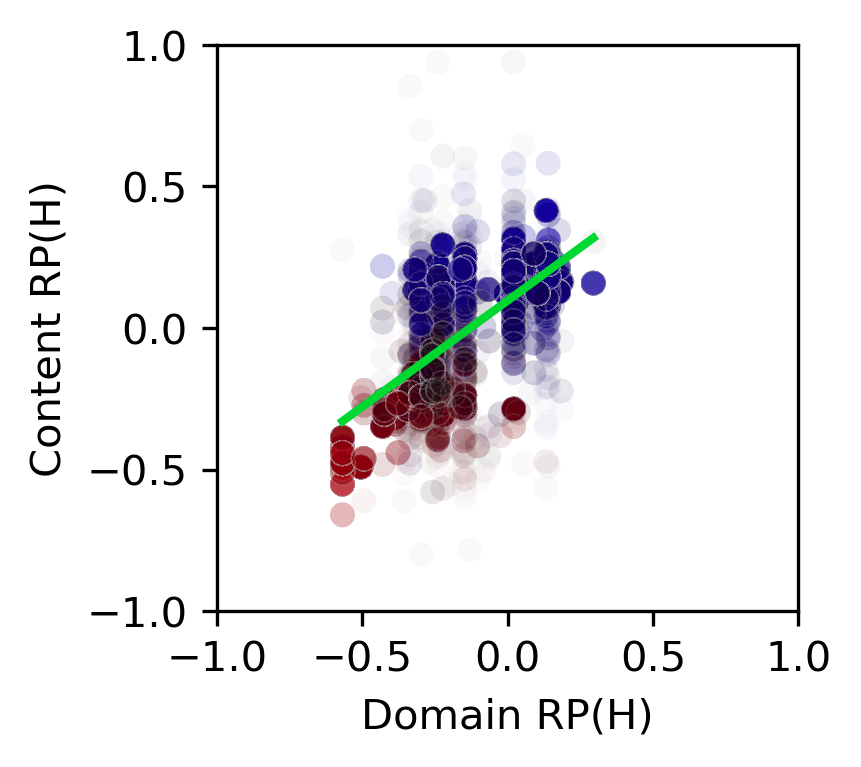"}
        \caption{Canada}
        \label{fig:domain_vs_content_scatter-can}
    \end{subfigure}
    \caption{Correlation between retweeted Domain RP(H) and Content RP(H).}
    \label{fig:domain_vs_content_scatter}
\end{figure}

\textbf{Domain RP(H) vs content RP(H):} When considering all retweets, a very strong positive correlation was identified in the Brazilian dataset with $r(23,639) = .895$*** and $\rho(23,639) = .900$***, and a strong positive correlation was identified in the Canadian dataset, with $(31,306) = .699$*** and $\rho(31,306) = .605$***. However, when neutral domains are analyzed in isolation, it is possible to observe a strong positive correlation between domain RP(H) and content RP(H), with $r(14,716) = .759$*** and $\rho(14,716) = .801$*** in the Brazilian dataset and a moderate positive correlation with $r(27,644) = .525$*** and $\rho(27,644) = .458$***, in the Canadian dataset. These correlations are significantly lower than between domain RP(H) and content RP(H) of polarized domains in Brazil, with $r(8,923) = .949$*** and $\rho(8,923) = .848$*** and slightly lower in the Canadian dataset, with $r(3,662) = .567$*** and $\rho(3,662) = .748$***.  Figure~\ref{fig:domain_vs_content_scatter} shows the scatter plot with the regression line (in green) for the correlation test involving domain RP(H), on the horizontal axis, and content RP(H) on the vertical axis, where each point is colour-coded considering the polarity value of users who made each retweet. In this figure, it is possible to observe what was identified in the correlation tests, which shows that neutral domains can generate engagement of users from both extremes of the political spectrum in a more balanced way, i.e., engage a similar proportion of users with different political orientations, potentially exposing those users to different points of view, thus, breaking the filter bubble. As we saw in the previous results, this balanced engagement does not mean that users with different political orientations engage with any content. In fact, the indication we have is that users are selective, engaging with content that matches their convictions.

\begin{figure}[ht!]
    \centering
    \begin{subfigure}{6cm}
        \centering
        \includegraphics[width=5cm]{"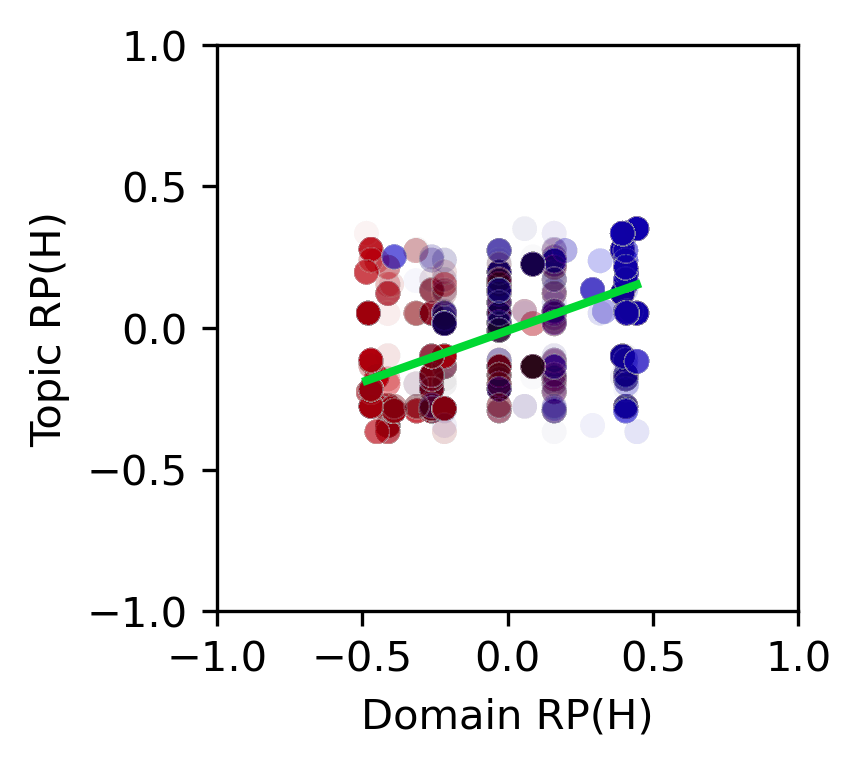"}
        \caption{Brazil}
        \label{fig:domain_vs_topic_scatter-bra}
    \end{subfigure}%
    \begin{subfigure}{6cm}
        \centering
        \includegraphics[width=5cm]{"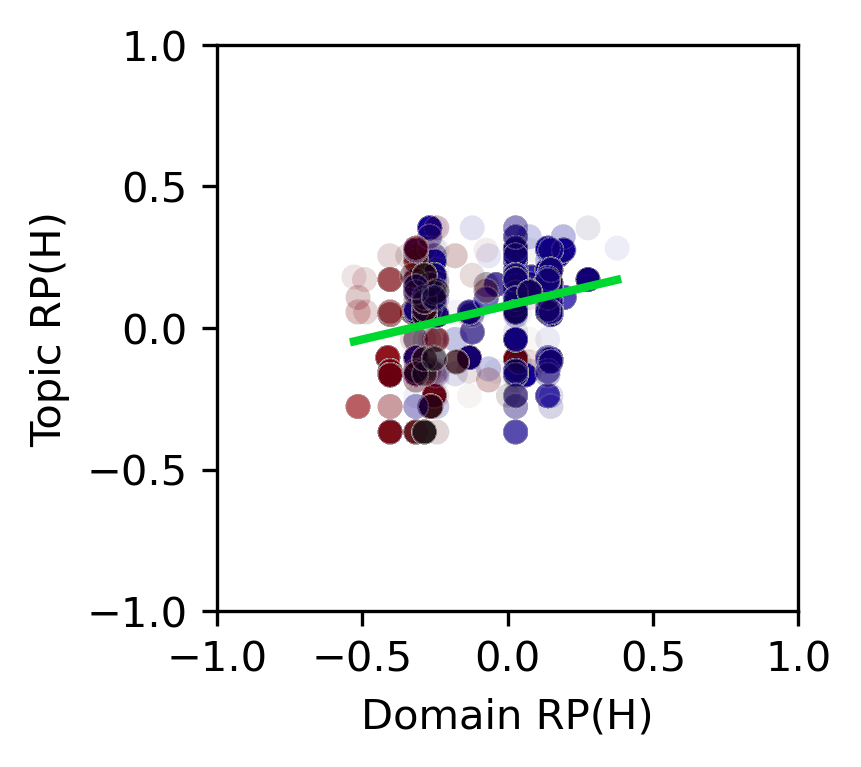"}
        \caption{Canada}
        \label{fig:domain_vs_topic_scatter-can}
    \end{subfigure}
    \caption{Correlation between retweeted Domain RP(H) and Topic RP(H).}
    \label{fig:domain_vs_topic_scatter}
\end{figure}

\textbf{Domain RP(H) vs topic RP(H):} Finally, in the test including all retweets involving the domain RP(H) and topic RP(H), a moderate positive correlation was identified in the Brazilian dataset, with $ r(19,029) = .548$*** and $\rho(19,029) = .538$***, and a weak positive correlation in the Canadian dataset, with $r(21,345) = .287$*** and $\rho(21,345) = .261$***, in the Canadian dataset. Figure~\ref{fig:domain_vs_topic_scatter} shows the scatter plot with the regression line (in green) for the correlation test involving domain RP(H), on the horizontal axis, and topic RP(H) on the vertical axis, where each point is colour-coded, considering the polarity value of users who made each retweet. Nevertheless, when we study neutral users in isolation, as it happened for previous tests, the correlation is slightly less intense when compared to polarized users. These results suggest that the political orientation of topics and domains is weakly related to retweeting behaviour. 

Overall, speaking to our fourth research question, we found that neutral domains tend to foster more engagement across partisan divisions than more partisan domains. Nevertheless, this high degree of cross-partisan exposure does not mean users from different partisan orientations engage with the same content to the same degree. Rather, they selectively engage with content that aligns with their partisan orientation. Moreover, this selective engagement is not based on the actual content but upon where it appears. Articles on similar topics generate very different responses depending on the media outlet that was dealing with it.

\section{Discussion and Conclusion}\label{sec:discussion}

The results indicate a common phenomenon across the Brazilian and Canadian datasets: neutral domains can distribute content that interests both sides of the political spectrum in a balanced way. Despite that, in most cases, users tend to engage only with content that reinforces their convictions, regardless of the topic. The same topic can generate different retweet behaviours, depending more strongly on the content or the domain’s political orientation (in polarized domains).

Social media platforms and mainstream news organizations have pursued initiatives to expose users to more frequently shared and neutral content, hoping that this effort would reduce polarization. The results presented in this study partially contrast with this received wisdom. We find that when users engage with neutral domains and are exposed to content that interests both sides of the political spectrum, they prefer to select the content that better fits their previous beliefs. In short, users seek to engage with information more aligned with their biases, even when exposed to contrasting information. Neither cross-cutting exposure nor ``neutral'' media alone seems to be enough to free people from their echo chambers. Intergroup contact with the same content may not have the expected effect in reducing polarization. Our results thus build on previous literature emphasizing the role of selective exposure in structuring online polarization \cite{iyengar2009red,Hameleers2018}. We offer greater confidence in the truth of these claims by using an updated centrality metric. This suggests selective exposure is not simply an artifact of prior measurement techniques but reflects something fundamental about engagement on social media.

On the other hand, evidence that neutral domains generate engagement from distinct polarized groups suggests news media can still function as a ``bubble reacher'', but in a different way. By evidencing prior biases as the decisive factor in building public opinion, selective exposure arguments tend to minimize the impact of news media. However, while news media may be ineffective in telling people what to think, others note it is more successful in telling people what to think about through ``agenda setting'' \cite{lippmann2017public,mccombs2020setting}. Based on that, news media can still promote cohesion by delineating what the important topics are, even when positions on these issues diverge. Our analysis speaks to this dynamic by finding that even when neutral domains are unable to upend partisan biases, they still manage to draw interest from across the political spectrum. Neutral domains can thus conceivably mediate polarization by integrating opposing partisans within common topical agendas, even if this means adopting antagonistic perspectives. Along similar lines, the fact that neutral domains can draw interest from both sides of the political spectrum could be interpreted as consistent with their promise to be impartial in regard to partisan  ``coverage bias'' \cite{d2000media}. Nowadays, this is an important point considering news media are constantly being attacked by country leaders who claim that they produce fake or biased news to weaken their power \cite{faris2017partisanship,huckfeldt_disagreement_2004}.

Despite the interesting results of this study, there are limitations that we should highlight. The use of trending hashtags for data collection could have impacted the retweet diversity and limited the content only to popular political subjects around election settings of each country. Because of that, the observed phenomena about the role of bubble reachers and how users behave when exposed to its content must not be interpreted as something universal for any polarized situation or online social network. Moreover, the action of retweeting was considered a sign of a user’s endorsement of a domain, content, or topic. However, retweets do not always indicate that a user is endorsing something. Regarding this point, users could even retweet content they don’t agree with to make fun of or mock it in front of their peers. In addition, the Twitter platform recommendation mechanisms may be generating filter bubbles \cite{pariser2011filter} through the selection of content published by traditional media that would potentially generate more engagement on either side of the political spectrum. Despite this, with the sample sizes of the analyzed datasets and the fact that a balance has been identified in the distribution of neutral media content among users in differing political spectrums, that possibility becomes less likely to have affected the obtained results. We recognize that the high granularity of the L, N, and R labels greatly simplified our analysis, however real extremists could behave differently compared to those who are merely biased, and those we considered neutral in their orientation were not necessarily so. Moreover, this categorization must be interpreted with care, since each situation has political characteristics far more complex than those expressed merely by L, N, or R labels, for example, users with the same political orientation could be acting as supporters for different political parties \cite{cochrane2015left}. Lastly, it is important to consider that the filtering of most active users on a weekly basis may have captured more politically motivated individuals for our analysis, which could inflate the polarization indicators -- they could be more extreme in their ideology and polarized. Despite this, there is a large number of users in the dataset, and politically motivated individuals could be in a smaller proportion when compared to individuals just sharing political content among other non-political content, even considering the election's setting.

For future studies, it is important to verify whether the same dynamic found in the engagement of users with content created by neutral top bubble reachers is replicated in other online social networks, such as Facebook, Whatsapp and Reddit or even in offline social networks. Cinelli et al. \cite{cinelli_2021}, for instance, find that platforms like Twitter which are organized around social networks and news feed algorithms generate greater homophilic clustering than others. Understanding this phenomenon in distinct contexts is important for consolidating policies or building mechanisms that aim to reduce polarization, whether online or offline \cite{stray_2021}. Such research could build on recent work outlining the different functions of cross-partisan brokerage \cite{keuchenius2021important,brocic2021influence} by elaborating the role of neutral bubble reachers. Drawing on our earlier discussion, this could mean exploring their role in topical ``agenda-setting''.

Regarding the role of news media, an important step is to assess the degree to which this media is able to distribute content in a balanced way to different social groups and how individuals behave when exposed to such content, not only considering the sharing behaviour, as performed in this work, but also in the discussion cascades generated from the consumption of online contents. In this sense, recent work on message cascades exchanged by Whatsapp \cite{caetano2019characterizing} users identified that the reach of messages with false political information was greater than that of other types of information. Similar research could be built in order to understand the user’s reaction through discussions, when exposed to information shared by news media, considering the political bias of users and the kind of information to which they were exposed.

It's also important to point out a possible extension for the \textit{intergroup bridging} algorithm: considering that the main purpose of the algorithm was to find nodes that could reach nodes with distinct labels from itself more efficiently, its mechanism could potentially be updated to count only shortest paths between nodes with different labels (political orientation, in the case of this research) while maintaining the same dynamic of the \textit{intergroup bridging} algorithm to minimize the effect of star nodes, as it inherited from bridgeness centrality. This is specially useful in problems involving finding efficient nodes that act as super-spreaders between distinct groups.

\begin{backmatter}

\section*{Acknowledgements}
The authors would like to thank Michelle Reddy for the ongoing discussions on the thematic of this study.

\section*{Funding}
All stages of this study was financed in part by CAPES - Finance Code 001,  project GoodWeb (Grant 2018/23011-1 from S\~{a}o Paulo Research Foundation - FAPESP), CNPq (grant 310998/2020-4), and by a Connaught Global Challenge Award.

\section*{Availability of data and materials}
The datasets generated and analyzed during the current study will be made available in a public repository after review.

  
\section*{Competing interests}
The authors declare that they have no competing interests.

\section*{Author's contributions}
JK ran the analysis and wrote the paper. TS, MB and DS conceptualized the analysis and wrote the paper. AG revised the paper. All authors read and approved the final manuscript.



\bibliographystyle{bmc-mathphys} 
\bibliography{bmc_article}      








\begin{table}[ht!]
    \caption{Heatmap data references for Figure~\ref{fig:content_heatmap}}
    \label{tab:content_heatmap}
    \begin{tabularx}{\textwidth}{lp{10cm}p{1cm}}
        \hline
        ID & Content URL for (a) Brazil &  RP(H) \\
        \hline
        1 &                            \url{https://jornalistaslivres.org/urgente-tse-recebe-acao-para-investigar-caixa-dois-de-bolsonaro-e-impugnar-sua-candidatura} &  -0.70 \\
        2 &   \url{https://noticias.uol.com.br/politica/eleicoes/2018/noticias/2018/10/18/medico-diz-que-bolsonaro-evolui-bem-mas-nao-cita-liberacao-para-debates.htm} &  -0.68 \\
        3 &                                                                      \url{https://lula.com.br/arrancada-haddad-cresce-6-pontos-em-minas-e-3-em-sao-paulo} &  -0.64 \\
        4 &                                                                                                                   \url{https://www.valor.com.br/u/5920771} &  -0.44 \\
        5 &                                                                       \url{https://renovamidia.com.br/bolsonaro-na-record-durante-o-debate-da-rede-globo} &   0.49 \\
        6 &                                        \url{http://www.caneta.org/checagens/checamos-haddad-mente-sobre-pedido-de-prisao-de-empresarios-sem-investigacao} &   0.51 \\
        7 &                                                                                                               \url{https://www.oantagonista.com/?p=122290} &   0.65 \\
        8 &                                                                                                               \url{https://www.oantagonista.com/?p=123030} &   0.66 \\
        9 &                                                             \url{https://istoe.com.br/bolsonaro-recebe-6138-dos-votos-validos-em-londres-haddad-tem-3844} &   0.79 \\
        10 &  \url{https://politica.estadao.com.br/noticias/eleicoes,mourao-diz-que-vai-processar-compositor-por-fala-sobre-tortura-azevedo-pede-desculpas,70002559739} &   0.82 \\
        \hline
        ID & Content URL for (b) Canada  &  RP(H) \\
        \hline
        1 &                                                                                  \url{https://www.ctvnews.ca/video?playlistid=1.4634949} &  -0.59 \\
        2 &                                                        \url{https://www.cp24.com/video?clipid=1807978\&binid=1.1127680\&playlistpagenum=1} &  -0.49 \\
        3 &                                \url{https://torontosun.com/opinion/columnists/jim-warren-justin-trudeau-deserves-another-shot-heres-why} &  -0.47 \\
        4 &  \url{https://press-presse.liberal.ca/fact-check/conservative-platform-includes-53-billion-in-cuts-including-14-billion-in-hidden-cuts} &  -0.45 \\
        5 &               \url{https://www.theglobeandmail.com/politics/article-liberal-mp-involved-in-cannabis-start-up-did-not-publicly-disclose} &   0.32 \\
        6 &                                                                                         \url{https://www.conservative.ca/advance-polls} &   0.37 \\
        7 &                             \url{https://montrealgazette.com/opinion/editorials/our-endorsement-best-choice-for-canada-is-andrew-scheer} &   0.42 \\
        8 &                           \url{https://tnc.news/2019/10/06/malcolm-trudeaus-so-called-carbon-offsets-bought-from-liberal-friendly-firm} &   0.46 \\
        9 &                                                                                                              \url{http://cpcp.cc/ghk-ry} &   0.59 \\
        10 &                              \url{https://election.ctvnews.ca/truth-tracker-are-bots-amplifying-trudeaumustgo-twitter-says-no-1.4612390} &   0.66 \\
    \end{tabularx}
\end{table}

\begin{table}[ht!]
    \caption{Heatmap data references for Figure~\ref{fig:topic_heatmap}}
    \label{tab:topic_heatmap}
    \begin{tabularx}{\textwidth}{lp{10cm}p{1cm}}
        \hline         
        ID & Topic keywords for (a) Brazil  &  RP(H)\\
        \hline
  1 &                       lutar, percentual, filho, levantamento, crescer, movimentar, voto\_valido, manha, cenario, jovem &  -0.44 \\
  2 &  battisti, italia, marinha, extraditar, ambientar, audio, reporter, paulo\_guedes, forcas\_armadas, defender\_democracia &  -0.42 \\
  3 &                        manifestacao, jornal, apoiadores, tv, aplicativo, controlo, combater, rio, domingo\_foto, tuite &  -0.36 \\
  4 &                           tratamento, saudar, escola, vidar, previdencia, comprar, crescer, globo, ampliar, pretender &  -0.27 \\
  5 &             londres, votacao, receber\_voto\_valido, perguntar, aprender, escola, acao, sexo, descontar, comprar\_pacote &  -0.14 \\
  6 &                     obrar, ex\_presidente, petrobras, depoimento, acusar, preso, site, bahia, policia\_federal, indicio &   0.05 \\
  7 &                                 conteudo, grupo, fux, agencia, censurar, enviar, servico, kit\_gay, apoiadores, ilegal &   0.07 \\
  8 &      torturar, mourao, geraldo\_azevedo, torturador, presar, hamilton\_mourao, show, artista, planar, mourao\_torturador &   0.23 \\
  9 &                   medico, hospital, paciente, receitar, rasgar\_receitar, unidade, giselda\_trigueiro, jose, uol, natal &   0.24 \\
 10 &                           evangelico, igreja, lideranca, pastor, cristao, deus, religioso, urna, vieira, confederacao &   0.28 \\\\
        
        \hline
        ID & Topic keywords for (b) Canada  &  RP(H) \\
        \hline 
  1 &                      rcmp, girl, snc\_lavalin, mr\_scheer, information, rebel, mr\_trudeau, levant, witness, energy &  -0.17 \\
  2 &                                   coalition, win\_seat, al, bloc, windsor, abortion, new\_democrats, guy, winnipeg &  -0.09 \\
  3 &                       genocide, apologize, bloc, apology, inquiry, blanchet, indigenous, racism, black, quebecer &  -0.07 \\
  4 &                                pipeline, protest, library, obama, oil, plane, hypocrisy, climate, rumour, murphy &   0.15 \\
  5 &                                                                       de, le, la, et, un, du, les, pour, que, au &   0.15 \\
  6 &                      elections\_canada, saroya, refugee, ballot, voting, racism, young, mr\_trudeau, apology, form &   0.16 \\
  7 &  infrastructure, spending, new\_democrats, mr\_trudeau, propose, indigenous, mr\_scheer, health\_care, deficit, gain &   0.33 \\
  8 &                andrew, protest, feeling, thinking, speech, cabinet, deserve\_elect, editorial, date, young\_people &   0.36 \\
  9 &                       bernier, beauce, abortion, worker, maxime\_bernier, bloc, levesque, director, present, farm &   0.39 \\
 10 &                            coal, water, health, china, harper, lean, environmental, mr\_harper, cap\_trade, mining &   0.47 \\
    \end{tabularx}
\end{table}


\end{backmatter}
\end{document}